\documentclass[sigconf]{acmart}

\AtBeginDocument{%
  }

\usepackage{longtable, multirow, booktabs, array, caption, amsmath, tabularx, enumitem, makecell}

\copyrightyear{2026}
\acmYear{2026}
\setcopyright{cc}
\setcctype{by-nc-nd}
\acmConference[CHI '26]{Proceedings of the 2026 CHI Conference on Human Factors in Computing Systems}{April 13--17, 2026}{Barcelona, Spain}
\acmBooktitle{Proceedings of the 2026 CHI Conference on Human Factors in Computing Systems (CHI '26), April 13--17, 2026, Barcelona, Spain}
\acmPrice{}
\acmDOI{10.1145/3772318.3790752}
\acmISBN{979-8-4007-2278-3/2026/04}

\begin{document}

\title[From Understanding Challenges in Patient Journeys to Deriving Design Implications for North Korean Defectors' Adaptation]{``I Should Know, But I Dare Not Ask'': From Understanding Challenges in Patient Journeys to Deriving Design Implications for North Korean Defectors' Adaptation}

\author{Hyungwoo Song}
\orcid{0009-0008-6649-9453}
\affiliation{%
  \department{Department of Intelligence and Information}
  \institution{Seoul National University}
  \city{Seoul}
  \country{Republic of Korea}
}
\email{rotto95@snu.ac.kr}

\author{Jeongha Kim}
\orcid{0009-0005-5660-8810}
\affiliation{%
  \department{College of Nursing}
  \institution{Seoul National University}
  \city{Seoul}
  \country{Republic of Korea}}
\email{jhakim@snu.ac.kr}

\author{Minju Kim}
\orcid{0009-0002-1957-7611}
\affiliation{%
  \department{College of Nursing}
  \institution{Seoul National University}
  \city{Seoul}
  \country{Republic of Korea}}
\email{lialos@snu.ac.kr}

\author{Duhyung Kwak}
\orcid{0009-0001-2512-0055}
\affiliation{%
  \department{College of Nursing}
  \institution{Seoul National University}
  \city{Seoul}
  \country{Republic of Korea}}
\email{rvanpersie20@snu.ac.kr}

\author{Minjeong Shin}
\orcid{0009-0003-1235-3436}
\affiliation{%
  \department{Department of Intelligence and Information}
  \institution{Seoul National University}
  \city{Seoul}
  \country{Republic of Korea}}
\email{shinmj1024@snu.ac.kr}

\author{Bongwon Suh}
\authornote{Co-corresponding authors.}
\orcid{0000-0001-5610-9265}
\affiliation{%
  \department{Department of Intelligence and Information}
  \institution{Seoul National University}
  \city{Seoul}
  \country{Republic of Korea}}
\email{bongwon@snu.ac.kr}

\author{Hyunggu Jung}
\authornotemark[1]
\orcid{0000-0002-2967-4370}
\affiliation{%
   \department{College of Nursing}
  \institution{Seoul National University}
  \city{Seoul}
  \country{Republic of Korea}}
\email{hyunggu@snu.ac.kr}

\renewcommand{\shortauthors}{Song et al.}

\begin{abstract}
  While it is known that North Korean defectors (NKDs) struggle with South Korea's healthcare system, the specific challenges of their patient journey remain underexplored.
  To investigate this, we conducted interviews with 10 NKDs about an 8-step patient journey and identified the clinical consultation step as a critical barrier for all participants, marked by three key challenges: expressing symptoms, managing social and cultural concerns, and overcoming language differences.
  In response, we developed Medibridge, a mobile prototype that allows users to rehearse with an AI doctor before a real hospital visit to generate a tangible ``Helper Note'' for their actual consultation.
  Our evaluation with 15 NKDs showed improvements in perceived communication capability, including greater expression clarity, reduced social and cultural concerns, and enhanced linguistic confidence.
  Our contributions include an empirical understanding of NKDs' healthcare challenges, a novel AI-powered rehearsal system that prepares users for real-world clinical communication, and design implications for inclusive technologies for displaced populations.
\end{abstract}



\begin{teaserfigure}
\centering
  \includegraphics[width=0.9\textwidth]{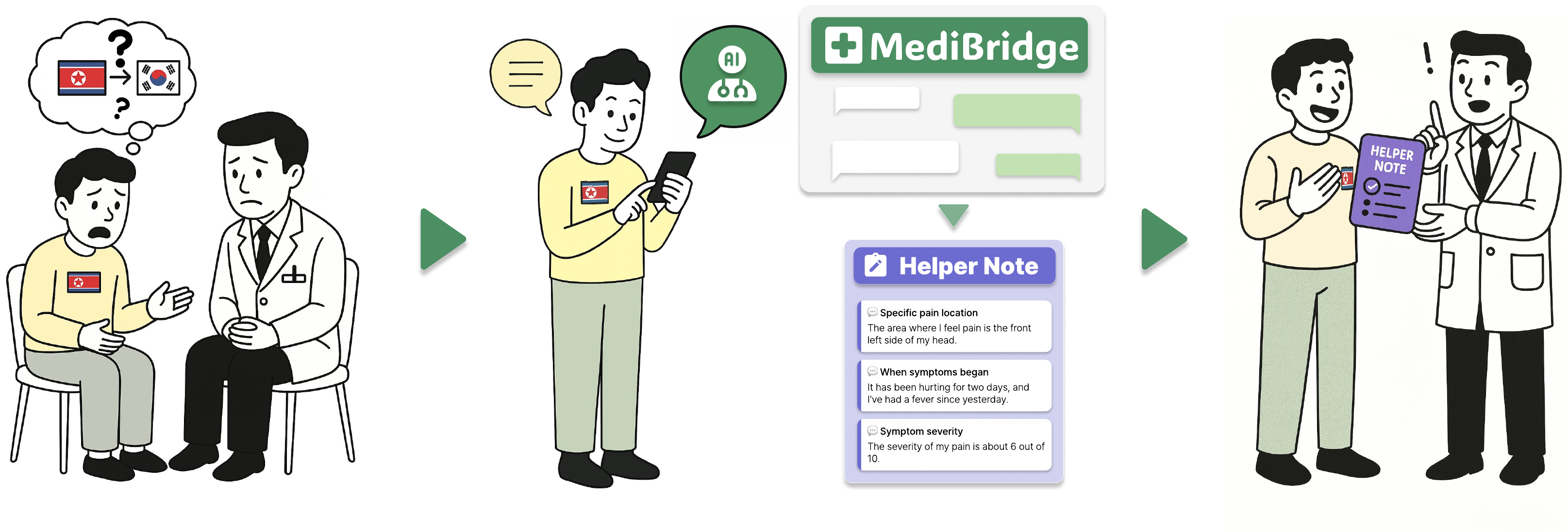}
  \caption{MediBridge supports healthcare communication for North Korean defectors through structured preparation. (Left) Communication challenges during medical consultations. (Center) AI-mediated rehearsal via a mobile interface. (Right) A structured ``Helper Note'' supporting patient–doctor interaction.}
  \Description{The figure consists of a left-to-right flow diagram with three main sections. The left section depicts a formative interview process covering eight steps of the patient journey, highlighted to identify challenges in the clinical consultation stage. The center section shows the MediBridge system, represented as a mobile interface enabling real-time dialogue practice with an AI doctor. The right section illustrates a user evaluation study comparing MediBridge with a simple version through experiments, surveys, and interviews, leading to summarized design implications for marginalized populations.}
  \label{fig:teaser}
\end{teaserfigure}

\maketitle

\section{Introduction}

The number of North Korean defectors (NKDs) entering South Korea has steadily increased since the 2000s, reaching a cumulative total of 34,314 as of 2025~\cite{MOU2025}.
They encounter various adaptation challenges during resettlement, including health disparities~\cite{Wee2024, Ahn2015}, cultural differences~\cite{Lee2022}, and economic hardships~\cite{Lee2015}.
Of the difficulties arising from systemic differences, the healthcare system represents a critical domain directly linked to individual survival.
While North Korea theoretically provides universal healthcare~\cite{OECD2020}, only 55.1\% of respondents reported receiving medical care for recent illnesses~\cite{lee2020health}.
Furthermore, North Korea lacks a modern digitized healthcare infrastructure, leaving NKDs unfamiliar with digital technologies integral to South Korean healthcare delivery.

South Korea provides comprehensive medical services through the National Health Insurance Service (NHIS)~\cite{Lee2020}, covering low-income populations and patients with chronic conditions.
However, NKDs face challenges in understanding this complex healthcare system, as effective utilization typically requires lifelong familiarity.
NKDs receive basic settlement education during their three-month program at Hanawon, the government support facility~\cite{RIR2023}.
However, this period proves insufficient for developing healthcare system literacy.
These difficulties are compounded by the NHIS's reliance on digital platforms for appointment scheduling and health record management~\cite{Lee2020, Noh2024}.

Recent HCI research explores NKDs' digital technology adaptation during resettlement.
Notably, Noh et al.~\cite{Noh2024} analyzed the impact of social media on identity reconstruction, while their subsequent co-design study~\cite{Noh2025} proposed twelve technology concepts addressing prejudice and disconnection.
While these studies illuminate the role of digital technology in general adaptation, healthcare-specific challenges remain underexplored.
Moreover, existing tools for migrant healthcare adopt generalized approaches~\cite{Freeman2013, Kisa2025, Narang2019}.
Most lack the behavioral scaffolding to overcome communication anxiety and deep-seated reluctance to engage with authority.
Consequently, current frameworks may not fully address circumstances particular to NKDs, such as permanent separation from homeland, linguistic barriers despite shared Korean heritage, and psychological patterns shaped by information-controlled environments.

\begin{figure*}[ht]
  \includegraphics[width=\textwidth]{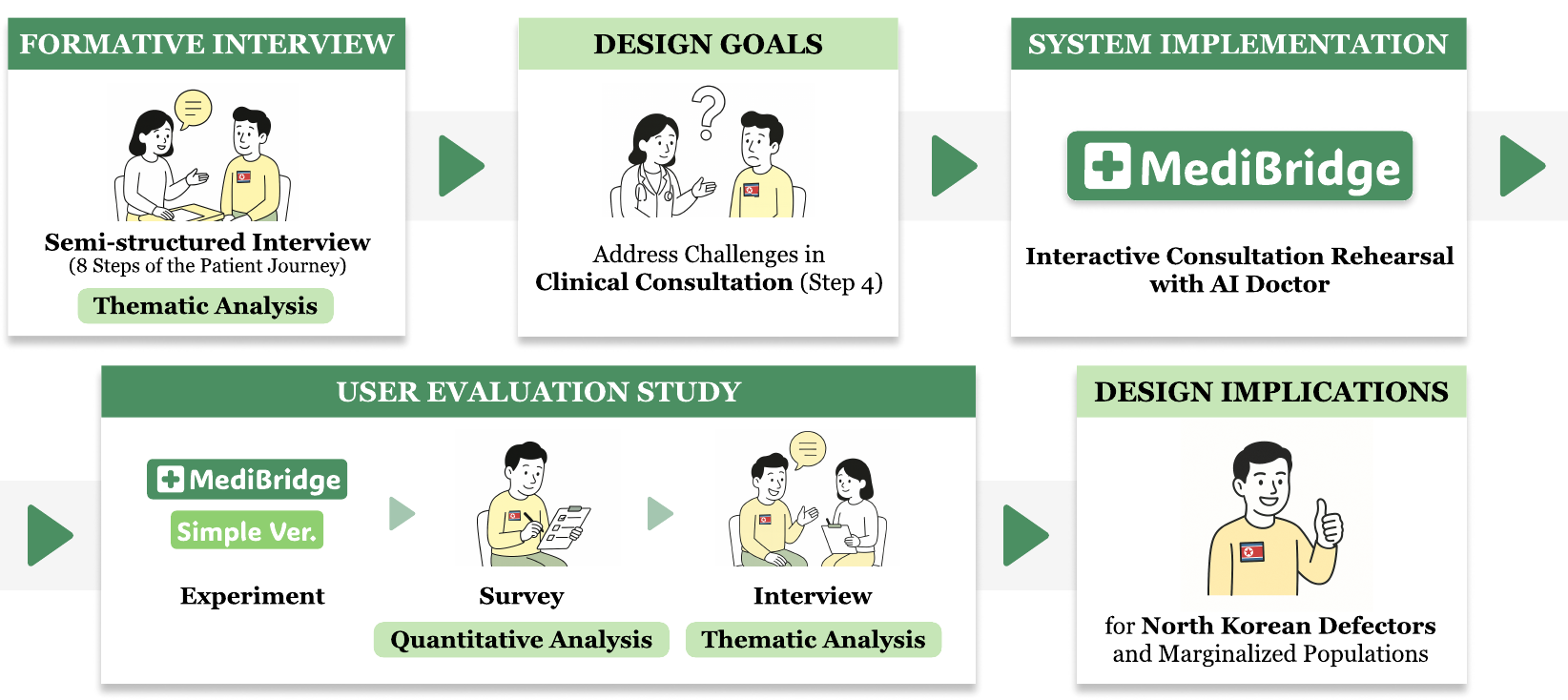}
  \caption{Research methodology from formative study to design implications. The study progresses through formative interviews identifying patient journey challenges, design goal development targeting clinical consultation difficulties, MediBridge system implementation with AI-powered dialogue and ``Helper Note'' generation, and three-condition evaluation study (pre-test baseline, Simple Version, MediBridge) leading to design implications for marginalized populations.}
  \Description{The figure is organized into two horizontal sections connected by directional arrows. The upper section presents the research and design process, beginning with semi-structured formative interviews across eight steps of the patient journey, followed by the identification of design goals focused on clinical consultation, and the implementation of the MediBridge system as a real-time AI doctor dialogue interface. The lower section depicts the user evaluation study, showing a comparative experimental setup between MediBridge and a simple version, followed by surveys and interviews analyzed using quantitative and thematic methods. Arrows indicate the sequential progression of research stages from left to right.}
  \label{fig:research_overview}
\end{figure*}

To investigate what specific challenges NKDs encounter in healthcare and where intervention is most needed, we conducted formative interviews with 10 NKDs exploring their experiences across an 8-step patient journey (see Fig.~\ref{fig:research_overview}).
Through thematic analysis, we identified the Clinical Consultation (step 4) as the critical barrier, characterized by three challenges: expressing symptoms, managing social and cultural concerns, and overcoming language differences.
In response, we developed MediBridge, a mobile prototype that enables users to rehearse medical conversations with a persona of an AI doctor and generates personalized ``Helper Note'' for actual clinical visits.
We evaluated the system with 15 NKDs through a comparative study across three conditions: pre-test baseline, a Simple Version with minimal functionality, and MediBridge, assessing perceived communication capability across six dimensions.

Our quantitative analysis revealed that MediBridge achieved higher perceived communication capability scores than the baseline across all six dimensions and showed higher scores than the Simple Version in essential information identification and social bias reduction.
Qualitative findings further showed patterns that may reflect NKD's unique circumstances.
Participants exhibited ambivalence toward their dialect—expressing desires to both preserve and modify it depending on context.
Notably, difficulties stemmed not from lack of information but from reluctance to question authority figures.
Participants also engaged readily with AI framed as a supportive doctor while resisting evaluative metrics—a distinction likely shaped by past experiences.
Collectively, the results suggest that healthcare technologies for populations with complex migration histories may benefit from addressing psychological dimensions alongside functional barriers.

This work offers four contributions.
First, we provide empirical understanding of NKDs' healthcare challenges through systematic investigation of their patient journey, identifying Clinical Consultation as the critical intervention point.
Second, we present MediBridge, a mobile prototype combining AI-powered rehearsal with personalized ``Helper Note'' generation to prepare users for real-world clinical encounters.
Third, through comparative evaluation with 15 NKDs, we provide evidence that the system improves perceived communication capability and reduces social and cultural concerns.
Finally, we derive design implications for healthcare technologies serving populations with complex migration histories, including considerations for navigating identity tensions, supporting agency development in users from authority-sensitive backgrounds, and designing feedback mechanisms that provide guidance without triggering evaluation anxiety.

\section{Background and Related Work}

This section situates our study by combining contextual background on North Korean defectors (NKDs) with a focused review of prior research on migrant healthcare challenges and technological interventions.

\subsection{Background} \label{sec:background}

We begin by establishing the sociopolitical, healthcare, and migration-specific conditions that shape North Korean defectors' healthcare experiences in South Korea.

North Korea operates as a totalitarian state under the Workers' Party's absolute control, fundamentally shaping citizens' lived experiences~\cite{KINU2018}.
This closed system has produced a population with severely limited exposure to market economies~\cite{OECD2020}, democratic institutions~\cite{KINU2021}, or concepts of individual rights~\cite{HRW2024, OHCHR2024}.
Citizens' digital access is tightly restricted: internet access rates approach nearly 0\%, approximately six million mobile phones in circulation as of 2020 cannot access international internet~\cite{Yoon2020, choi2022comparative}, and the government mandates surveillance software on all devices~\cite{williams2019digital}.
This extensive isolation creates a population with minimal familiarity with the digital infrastructure, consumer choice, and individual agency that characterize modern South Korean society.

While North Korea constitutionally guarantees ``complete free medical care'' for all citizens, the actual healthcare system fails to deliver such promises.
A comprehensive 2014--2015 survey of 383 NKDs found that only 55.1\% received medical services when ill~\cite{LeeHayoung2020}.
Most hospitals were built in the 1960s--70s and lack reliable electricity, water, and heating systems~\cite{Toimela2017}, and only 10\% of medications are provided through official channels, while 60.5\% must be purchased from pharmacies and 42.5\% from street vendors~\cite{LeeHayoung2020}.
This shadow healthcare economy fosters a culture of self-treatment and widespread distrust of official medical institutions, with initial health assessments upon arrival in South Korea revealing an average of 7.8 concurrent medical conditions per person~\cite{Lee2020}.

Most NKDs (57\%) are aged 20--39, representing prime working-age adults raised within North Korea's isolated system.
The defection process itself is extraordinarily dangerous and protracted, typically spanning 3--4 years and involving illegal border crossing, prolonged hiding, and perilous routes through multiple countries~\cite{ICG2006, Han2020, KINU2003}.
Such a trajectory subjects defectors to additional trauma and adaptation challenges beyond their initial North Korean experiences.

Upon arrival in South Korea, all NKDs undergo mandatory three-month residence at Hanawon, the government-operated settlement support center, receiving 400 hours of intensive education covering social adaptation, vocational training, and digital literacy~\cite{MOU2022, RIR2023}.
However, such programs face structural limitations in compressing decades of missed technological and social development into 12 weeks~\cite{CrossingBorders2020, Noh2024, MOU2025}.
In particular, the curriculum primarily addresses administrative procedures for medical aid applications while providing limited guidance on appointment scheduling, insurance navigation, and effective communication with healthcare providers~\cite{Yoon2007}.
As a result, many NKDs enter South Korea's complex healthcare system without adequate preparation, facing difficulties that persist well beyond the initial settlement period~\cite{Min2008, Kim2021}.

These persistent difficulties cannot be attributed solely to educational shortcomings; they stem from the fundamentally distinct circumstances of NKDs as a migration population~\cite{Poorman2019}.
Unlike economic migrants who maintain homeland connections or refugees who may eventually repatriate, NKDs face permanent and irreversible separation from their country of origin, families, and cultural systems~\cite{KINU2022, MOU2023}.
NKDs receive automatic citizenship~\cite{Poorman2019, Song2018} and Medical Aid Type 1 coverage for five years~\cite{ROK_Law2007, KimKeunA2021}, institutional arrangements that may suggest relatively smooth integration.
At the same time, their shared ethnic and linguistic heritage with South Koreans often generates expectations of seamless social integration, which can obscure more than 70 years of divergent social, political, and institutional development~\cite{Denney2024, Wee2024}.
This divergence manifests concretely in healthcare contexts: linguistic analysis reveals only 44\% similarity in dictionary terms between North and South Korean usage~\cite{Lee2021}, NKDs must transition from a centralized healthcare system to an insurance-based system emphasizing patient choice~\cite{Joo2020}, and social stigmatization combined with fear of identity disclosure further impedes healthcare utilization~\cite{Chun2022, LeeDohhee2020, Ahn2015, Kim2019}.
These converging linguistic, systemic, and psychological barriers constitute a distinct set of healthcare challenges that existing integration frameworks have yet to adequately address.

\subsection{Related Work}

Building on the contextual foundation established above, we review existing research on migrant healthcare challenges and technological interventions, and identify key conceptual and empirical gaps that motivate the present study.

\subsubsection{\textbf{General Challenges Faced by Migrants in Healthcare Systems}}

Migrants face substantial barriers when navigating host-country healthcare, driven by language, cultural unfamiliarity, and socioeconomic constraints~\cite{Omeri2006, Espinoza2014, Hargreaves2006, Ahmed2016}.
Language barriers are the most pervasive, as migrants must convey complex medical concerns across different linguistic systems, often without professional interpreters~\cite{VanSon2013, Yelland2014}.
Cultural unfamiliarity with local healthcare system further compounds the situation; migrants frequently lack knowledge of appointment procedures, insurance requirements, and referral processes, and recent reviews indicate that such challenges are exacerbated by organizational and administrative complexities inherent in host countries' health policies~\cite{Partyka2024}.
This unfamiliarity often results in overreliance on emergency medical services for primary care needs~\cite{Hargreaves2006}.
Economic constraints create additional access barriers, as many migrants hold low-wage jobs with inflexible schedules and minimal health benefits~\cite{Ahmed2016}.
Furthermore, some migrants may prefer traditional medicine from their home countries, creating potential conflicts with Western medical approaches, particularly when coupled  with mistrust toward unfamiliar healthcare systems~\cite{VanSon2013}.

\subsubsection{\textbf{Existing Tools for Healthcare System Adaptation}}

Current approaches to supporting migrant healthcare adaptation primarily focus on three areas: communication improvement, accessibility enhancement, and health information navigation.
Communication tools include digital storytelling platforms for health education~\cite{Kisa2025}, mobile translation applications providing one-click phone interpreter services~\cite{Narang2019}, and simulation-based training programs for healthcare providers~\cite{Skjerve2023}.
Accessibility improvements encompass culturally sensitive AI-powered chatbots for specific health topics (such as HPV vaccination for Korean Americans)~\cite{Kim2025} and culturally adapted mobile health applications translated for specific ethnic communities~\cite{Castillo2022}.
Health navigation tools range from comprehensive frameworks designed to remove barriers for minority groups~\cite{Freeman2013} to validated healthcare navigation capacity scales that assess migrants' ability to effectively utilize healthcare systems~\cite{Yeo2025}.
However, the tools demonstrate significant limitations: most are developed for general migrant populations without addressing the distinctive circumstances of popluations from authoritarian systems, such as the linguistic, psychological, and systemic barriers documented in the Section~\ref{sec:background}.
Moreover, evaluation of actual effectiveness remains limited with few studies conducting comprehensive usability assessments, and notably, no tools have been specifically developed and tested within actual healthcare settings for real-world implementation.

\subsubsection{\textbf{HCI Research on North Korean Defectors}}

The HCI community has shown limited engagement with North Korean defectors, with only recently emerging empirical research examining their unique experiences~\cite{Noh2024, Noh2025}.
Recent work by Noh et al.~\cite{Noh2024} represents the first comprehensive HCI investigation into how NKDs use digital technology during their transition to South Korea's highly connected society, revealing through interviews with 21 defectors that social media serves as a ``double-edged sword'' that enables identity construction while exacerbating gaps between real and ideal selves.
Building upon these foundational insights, the same team used speculative co‑creation with 22 defectors to envision 13 technologies addressing identity stigma, disconnection from the past, and adaptation to digital society~\cite{Noh2025}.
Both studies also revealed considerable methodological challenges in conducting HCI research with this population, including privacy concerns, identity disclosure risks, and the need for specialized approaches that account for their unique circumstances.
Although these initial contributions provide valuable insights into the technological needs of NKDs, healthcare system navigation, a fundamental survival need, remains insufficiently addressed in HCI research.

\subsubsection{\textbf{Research Gaps and Study Objectives}}

The reviewed literature reveals three critical gaps that collectively limit our understanding and support of NKDs' healthcare adaptation. 
First, existing migrant healthcare tools adopt generalized approaches that fail to address the distinctive circumstances of NKDs, including permanent homeland separation, unique linguistic barriers, and psychological challenges arising from irreversible displacement.  
Second, although recent HCI work examines defectors' technology use during social adaptation, healthcare system navigation remains largely unaddressed in the field.
Third, current healthcare adaptation tools lack comprehensive evaluation of their real-world effectiveness, with most studies providing limited evidence of actual impact on users' healthcare experiences.

To address these gaps, this study investigates how technology can effectively support NKDs' healthcare communication challenges.
Specifically, we pursue three objectives:
(1) empirically investigate the specific healthcare challenges NKDs encounter throughout their patient journey to identify critical intervention points,
(2) design and evaluate a mobile prototype that addresses these specific communication challenges, and
(3) derive design guidelines for healthcare technologies serving populations with complex migration backgrounds.
Through systematic investigation of NKDs' healthcare experiences and iterative system development, this work aims to establish a foundation for more effective, culturally informed approaches to supporting vulnerable populations' healthcare access.

\section{Formative Interview: Understanding Patient Journey Challenges}

\begin{figure*}[t]
  \includegraphics[width=\textwidth]{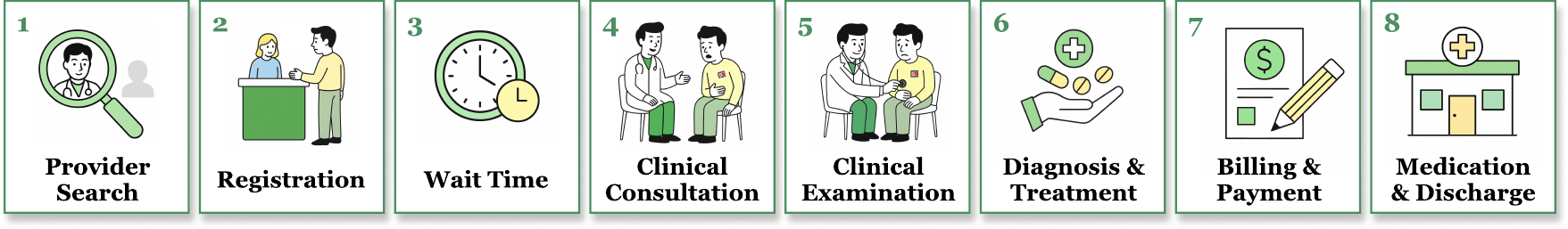}
  \caption{8-Step Patient Journey Framework. \textbf{(1) Provider Search}: finding appropriate healthcare providers, \textbf{(2) Registration}: completing intake procedures, \textbf{(3) Wait Time}: managing appointment scheduling and waiting periods, \textbf{(4) Clinical Consultation}: communicating symptoms and concerns with healthcare providers, \textbf{(5) Clinical Examination}: undergoing medical assessments, \textbf{(6) Diagnosis \& Treatment}: receiving medical decisions and treatment plans, \textbf{(7) Billing \& Payment}: handling financial transactions, and \textbf{(8) Medication \& Discharge}: completing prescriptions and follow-up instructions.}
  \Description{The figure presents a linear patient journey composed of eight sequential stages arranged from left to right. The stages include provider search, registration, wait time, clinical consultation, clinical examination, diagnosis and treatment, billing and payment, and medication and discharge. Each stage is represented by an icon and a label, indicating the progression of a patient through the healthcare system.}
  \label{fig:patient_journey}
\end{figure*}

To identify the critical barriers North Korean defectors encounter in South Korean healthcare, we conducted formative interviews exploring their experiences across the entire patient journey (see Fig.~\ref{fig:patient_journey}).
This formative phase aimed to pinpoint the most challenging touchpoints and derive user needs for system intervention.

\subsection{Method}

\subsubsection{\textbf{Participants}}

\begin{table*}[t]
\centering
\caption{%
Demographic Characteristics of Participants Across Both Studies. 
PID denotes participant ID. Study participation is marked with \checkmark{} (participated) or -- (not participated).
Age indicates age bands (e.g., 20--30), and Gender is noted as Female or Male.
Duration in South Korea indicates years of residence at the time of the study.
Transit Countries and Duration in Transit refer to the main country of residence and years spent there before arriving in South Korea.}
\Description{The table lists 16 participants (P1–P16) with information on study participation, age group, gender, duration of residence in South Korea, transit countries, and years spent in transit prior to arrival.
Ten participants took part in the formative interview study, and twelve participated in the evaluation study, with overlap across phases.
Participants vary in age from their 20s to 60s, include both male and female participants, and represent diverse migration trajectories primarily involving China, as well as other transit countries.
This demographic overview provides contextual grounding for interpreting participants’ healthcare experiences and system evaluation results.}

\label{tab:participant-info}
\begin{tabular}{c c c c c c c c}
\toprule
\textbf{PID} &
\textbf{\parbox[ht]{1.5cm}{\centering Formative Study}} &
\textbf{\parbox[ht]{1.5cm}{\centering Evaluation Study}} &
\textbf{Age} &
\textbf{Gender} &
\textbf{\parbox[ht]{2cm}{\centering Duration in South Korea\\(years)}} &
\textbf{\parbox[ht]{2cm}{\centering Transit Countries}} &
\textbf{\parbox[ht]{2cm}{\centering Duration in Transit\\(years)}} \\
\midrule
P1 & \checkmark & \checkmark & 50--60 & Male & 8 & Russia & 7 \\
P2 & \checkmark & \checkmark & 30--40 & Female & 6 & China & 6 \\
P3 & \checkmark & \checkmark & 30--40 & Female & 8 & China & 2 \\
P4 & \checkmark & \checkmark & 30--40 & Female & 7 & China & 10 \\
P5 & \checkmark & \checkmark & 30--40 & Female & 10 & China & 15 \\
P6 & \checkmark & -- & 20--30 & Male & 10 & China, Laos & 1 \\
P7 & \checkmark & \checkmark & 20--30 & Female & 10 & China, Thailand & 1 \\
P8 & \checkmark & \checkmark & 30--40 & Female & 9 & China & 4 \\
P9 & \checkmark & \checkmark & 20--30 & Female & 10 & China, Philippines & 1 \\
P10 & \checkmark & \checkmark & 20--30 & Male & 9 & China, Thailand & 1 \\
P11 & -- & \checkmark & 20--30 & Female & 9 & China, Laos & 1 \\
P12 & -- & \checkmark & 30--40 & Male & 12 & China & 1 \\
P13 & -- & \checkmark & 50--60 & Male & 23 & Singapore & 1 \\
P14 & -- & \checkmark & 50--60 & Female & 23 & Singapore & 1 \\
P15 & -- & \checkmark & 40--50 & Female & 17 & China & 10 \\
P16 & -- & \checkmark & 30--40 & Male & 23 & Singapore & 1 \\
\bottomrule
\end{tabular}
\end{table*}

We recruited 10 NKDs using a trust-based snowball sampling strategy through resettlement support networks (Table~\ref{tab:participant-info}).
This approach was essential to overcome the hidden population's reluctance to reveal identities due to stigma.
Inclusion criteria required participants to have:
(1) resided in South Korea for less than 10 years to ensure vivid recollections of adaptation,
(2) utilized healthcare facilities annually, and
(3) independently sought health information to ensure we recruited participants who had personally confronted systemic information barriers.

\subsubsection{\textbf{Procedure}}

Semi-structured interviews (60--90 minutes) were conducted in pairs by members of our research team.
Participants chose between in-person meetings and remote video conferencing based on their preference.
With explicit consent, sessions were audio-recorded and transcribed;
in one instance where recording was declined due to privacy concerns, detailed field notes were taken immediately.
Data collection and analysis were performed in Korean to capture linguistic nuances, with representative quotes translated into English for this manuscript.
Compensation of 20,000 KRW was provided to each participant, and all procedures for both the formative interview and the evaluation study were approved by the Institutional Review Board.

\subsubsection{\textbf{Researcher Positionality}}
Our research team consists of seven South Korean HCI researchers, and we acknowledge that inherent power dynamics may arise when studying the experiences of North Korean defectors.
To mitigate this positionality gap, one team member who has immediate family ties to the NKD community facilitated a community consultation process throughout both the formative interview and the evaluation study.
This team member's family member, a North Korean defector, served as a community consultant who advised on culturally sensitive aspects of the study design and interview protocol prior to data collection, including appropriate language, potential sources of discomfort, and framing of questions.
The finalized protocols were then applied consistently within each study phase.
The team member with NKD family ties also participated in a subset of interviews as one of the two interviewers, which further helped foster psychological safety for participants.
Throughout the analysis, the community consultant additionally reviewed our interpretations to reduce the risk of misrepresentation.

\subsubsection{\textbf{Interview Protocol}}

The protocol was structured around a deductive framework based on the eight-step patient journey framework~\cite{joseph2020patient} adapted for the South Korean context (see Fig.~\ref{fig:patient_journey}).
Within this predefined structure, we employed a \textit{what-why-how} questioning strategy~\cite{Kallio2016} to systematically examine each step.
This approach allowed us to elicit specific incidents (\textit{what}), identify the underlying causes of difficulties (\textit{why}), and understand participants' behavioral responses (\textit{how}) at every touchpoint.
The comprehensive interview guide used in this study is provided in Appendix A of the Supplementary Material for reproducibility.

\subsubsection{\textbf{Thematic Analysis \& Design Goal Formulation}}

We conducted thematic analysis~\cite{terry2017thematic} using a paired coding strategy to ensure rigor.
For each transcript, two researchers independently performed open coding, resolving discrepancies through discussion or third-party arbitration.
The consolidated codes underwent bottom-up hierarchical clustering: specific difficulties (Level 1) were grouped into intermediate themes (Level 2) and synthesized into overarching categories (Level 3) mapped to the patient journey steps, with the final structure validated by the full research team.

Following the analysis, the Clinical Consultation (Step 4) emerged as the critical intervention point, unanimously cited by all participants as a primary barrier and representing the core communicative interaction of healthcare seeking.
To translate this priority into design goals, four team members conducted an iterative process spanning two weeks (approximately 10 hours of collaborative sessions).
This process involved converting validated codes into ``User Needs'' statements through affinity diagramming and subsequently formulating ``How Might We'' (HMW) questions.
The final three design goals were established only after reaching unanimous consensus on these goals.

\subsection{Findings}

Our analysis revealed that North Korean defectors (NKDs) encounter systemic and psychological barriers throughout their entire patient journey.
However, the \textbf{Clinical Consultation (Step 4)} emerged as the universal and most critical bottleneck where communication breakdowns most severely impacted health outcomes.
Therefore, we structure our findings into two parts: first, a holistic overview of the systemic barriers across the journey (Steps 1--3, 5--8), and second, an in-depth analysis of the consultation phase (Step 4) which directly informed our design interventions.

\subsubsection{\textbf{Challenges Across the Patient Journey}}

\begin{table*}[ht]
\caption{Key Challenges Across the Patient Journey (Excluding Step 4). This table presents the primary systemic and psychological barriers identified in our analysis. Detailed coding structures supporting these findings are available in Appendix B.}
\Description{The table presents identified barriers across seven stages of the patient journey, excluding the clinical consultation stage, which is analyzed separately. For each step, the table summarizes recurring systemic challenges (e.g., institutional complexity, procedural unfamiliarity) and psychological challenges (e.g., anxiety, identity-related concerns) derived from thematic analysis of formative interviews. Barriers are organized by journey stage to illustrate how difficulties are distributed across the healthcare process rather than concentrated in a single point. The full coding structure supporting these summaries is provided in Appendix B.}
\label{tab:challenges_summary}
\centering
\small
\renewcommand{\arraystretch}{1.3}
\begin{tabularx}{\textwidth}{p{0.15\textwidth} X}
\toprule
\textbf{Step} & \textbf{Identified Barriers} \\
\midrule
\textbf{1. Provider Search} & 
• \textbf{Systemic Complexity:} Confusion caused by the unfamiliar hierarchy of primary, secondary, and tertiary hospitals. \newline
• \textbf{Information Reliability:} Difficulty establishing a baseline for trust when evaluating online reviews. \\
\hline

\textbf{2. Registration} & 
• \textbf{Protocol Unfamiliarity:} Logistical hurdles and confusion with kiosks and check-in procedures. \newline
• \textbf{Identity Anxiety:} Distinct fear of stigma when required to reveal NKD status or ID. \\
\hline

\textbf{3. Wait Time} & 
• \textbf{Process Uncertainty:} Nervousness about missing one's turn due to a lack of understanding of the queuing system. \newline
• \textbf{Environmental Maladaptation:} Feeling overwhelmed by the scale and crowding of large hospitals. \\
\hline

\textbf{5. Clinical Exam} & 
• \textbf{Information Void:} Passive confusion resulting from insufficient explanations about procedures. \newline
• \textbf{Wayfinding Issues:} Difficulty navigating complex hospital layouts to find examination rooms. \\
\hline

\textbf{6. Diagnosis \&\newline Treatment} & 
• \textbf{Power Dynamics:} Insufficient explanations forced patients into an uncomfortable, proactive role. \newline
• \textbf{Linguistic Divergence:} Struggles with NK-SK medical terminology differences. \\
\hline

\textbf{7. Billing \& Payment} & 
• \textbf{Fee Complexity:} Confusion regarding separated charges (consultation vs.\ medication). \newline
• \textbf{Behavioral Inertia:} Inability to compare costs effectively, discouraging provider switching. \\
\hline

\textbf{8. Medication \&\newline Discharge} & 
• \textbf{Information Asymmetry:} Lack of explanation regarding drug components and necessity. \newline
• \textbf{Valuation Gap:} Misunderstanding the value of prescribed medication, leading to non-adherence. \\
\bottomrule
\end{tabularx}
\end{table*}

Table~\ref{tab:challenges_summary} summarizes the primary barriers identified at each touchpoint (The complete thematic coding tree is provided in Appendix B of the Supplementary Material.)

To illustrate how these barriers manifested in participants' lived experiences, we highlight representative accounts from each step below.
In \textbf{Step 1 (Provider Search)}, participants struggled with both systemic structure and information reliability; P8 noted the hierarchical complexity: \textit{``Hospitals are divided into primary, secondary, and tertiary levels. Initially, I thought going to a big hospital would cure me faster.''}
In \textbf{Step 2 (Registration)}, identity exposure was a recurring concern, as P3 described: \textit{``When I talk to people, they notice that I am a North Korean defector. Even though they may not actually discriminate against me, I disliked the possibility that they might.''}
\textbf{Step 3 (Wait Time)} generated process-related anxiety; P7 recalled: \textit{``When I was suddenly called, I felt nervous and confused, wondering if it was my turn.''}

In \textbf{Step 5 (Clinical Examination)}, insufficient procedural explanations led to passive waiting, as P5 noted: \textit{``I didn't know where I should go, so I just sit and wait until someone call my name.''}
\textbf{Step 6 (Diagnosis \& Treatment)} revealed linguistic barriers distinct from general fluency; P5 shared: \textit{``Language is really difficult for me. I explained a symptom, but I thought, what was that term again? Medical symptom names don't come to mind.''}
In \textbf{Step 7 (Billing \& Payment)}, P4 described the compounding confusion: \textit{``Payment is difficult because sometimes there are separate charges for consultation fees and medication costs. But most of the time, they don't explain it kindly.''}
Finally, in \textbf{Step 8 (Medication \& Discharge)}, P1 highlighted the valuation gap: \textit{``If they told me this is a precious medicine and explained why I need to take it for three months, I wouldn't throw it away and would take it all.''}

\subsubsection{\textbf{Challenges and Derived Design Goals from Step 4: Clinical Consultation}}

The Clinical Consultation (Step 4) represents the critical interaction phase where patients communicate symptoms and receive medical guidance.
All ten participants identified this step as the most challenging aspect of their journey, reporting three categories of barriers that significantly hindered consultation effectiveness.
This section examines these challenges and derives the specific design goals that informed our system development.

\paragraph{\textbf{Ambiguity in Symptom Expression (Content).}}
All participants expressed uncertainty regarding \textit{what} information to share and \textit{how much} detail was appropriate, suggesting a lack of a clear mental model for medical communication in South Korea.
P8 highlighted the anxiety stemming from this ambiguity: \textit{``I worry I might miss out on the quality of care, because I don't know what to ask or tell the doctor.''}
Similarly, P4 noted confusion about specificity: \textit{``I'm not sure how detailed I should be when discussing personal symptoms.''}
To address this ambiguity, we derived \textbf{Design Goal 1: Structured Communication Support}.
This goal emphasizes that an effective intervention could provide a structured framework to facilitate systematic symptom organization, thereby enabling users to identify essential information and practice appropriate levels of detail.

\paragraph{\textbf{Social and Cultural Inhibition (Psychological Barrier).}}
Psychological barriers appeared to significantly impede honest self-disclosure and proactive questioning.
Participants reported a reluctance to reveal their background due to fear of judgment and a cultural tendency to avoid questioning authority figures.
P6 stated, \textit{``I can't honestly express my background··· worried about how they might judge me.''}
This was often compounded by a perceived hierarchy, as P5 noted: \textit{``I feel embarrassed to ask again, because questioning doctors feels inappropriate.''}
Furthermore, both P2 and P9 identified older male physicians as particularly intimidating, describing them as \textit{``paternalistic.''}
This pointed to the importance of \textbf{Design Goal 2: Psychological Scaffolding}.
We identified that a support system would need to go beyond functional training to include psychological preparation mechanisms, aiming to help users manage social concerns and build the confidence to exercise their rights as patients.

\paragraph{\textbf{Linguistic Anxiety (Communication).}}
Linguistic divergence appeared to create a bidirectional source of anxiety: uncertainty about using South Korean medical terminology and concern regarding the provider's comprehension of North Korean expressions.
P10 described the burden of adaptation: \textit{``I feel burdened about how to express my headaches in the South Korean way.''}
Conversely, P3 expressed fear of being misunderstood: \textit{``I worry that doctors won't understand when I use North Korean.''}
These findings informed \textbf{Design Goal 3: Adaptive Language Support)}.
This suggests that design solutions should likely offer adaptive mechanisms to bridge terminology gaps—aiming not merely for translation, but to strengthen communication confidence through practice in a linguistically safe environment.

\section{MediBridge}
Building on the three design goals derived from the formative study, we developed MediBridge, a mobile prototype that enables users to rehearse medical consultations through AI role-play and transforms their practice into a tangible ``Helper Note'' for use during actual clinical visits.
The system was implemented as a responsive web application with GPT-4o~\cite{openai2024gpt4o} as the core reasoning engine.
This section details how each design goal was operationalized into specific features and technical implementations, followed by a user scenario illustrating the complete experience.

\subsection{Design Goals to System Implementation}
We systematically translated the three Design Goals (DG) into system features, ensuring that each challenge identified in the formative study was addressed through specific functional and technical implementations.

\subsubsection{\textbf{DG 1: Structured Communication Support}}
To address the ambiguity in symptom expression—where participants lacked a clear mental model for \textit{what} to share and \textit{how much} detail was appropriate—MediBridge provides a guideline driven feedback loop comprising three integrated components.

First, \textbf{pre-practice guidelines} establish a clear mental model by presenting ten essential consultation elements derived from medical communication standards: five items to share (e.g., \textit{pain location}, \textit{current medications}, \textit{allergy history}) and five items to understand (e.g., \textit{diagnosis rationale}, \textit{potential side effects}).
Second, \textbf{interactive practice} enables iterative skill development through voice or text simulations with a persona of an AI doctor, allowing users to rehearse the flow of medical conversations.
Third, \textbf{criteria-based evaluation} provides specific assessments against the ten guidelines, offering actionable suggestions for missing or insufficiently detailed information.

The evaluation functionality analyzes conversation logs against the ten essential criteria using structured prompting enforcing strict JSON output.
This output populates visual evaluation cards with graded scores (Low/Mid/High) for each criterion, allowing users to identify gaps (e.g., receiving a 30\% score for \textit{Specific pain location} while achieving 70\% for \textit{Severity}) and receive suggestions such as ``Clearly describe where the pain is located.''

\subsubsection{\textbf{DG 2: Psychological Scaffolding}}
To mitigate the social and cultural inhibition that impeded honest self-disclosure, including fear of judgment and reluctance to question authority figures, the system integrates psychological safety mechanisms throughout the user flow.

First, \textbf{explicit rights education} frames proactive questioning as a legitimate right rather than a burden, countering cultural deference.
Second, \textbf{safe exposure} through an intentionally challenging persona of an AI allows users to practice assertiveness and experience authority-figure interactions without real-world consequences.
Third, \textbf{positive reinforcement algorithms} highlight communication strengths to build self-efficacy and validate users' natural expression patterns, rather than focusing solely on deficits.

\begin{figure*}[t]
  \centering
  \includegraphics[width=\linewidth]{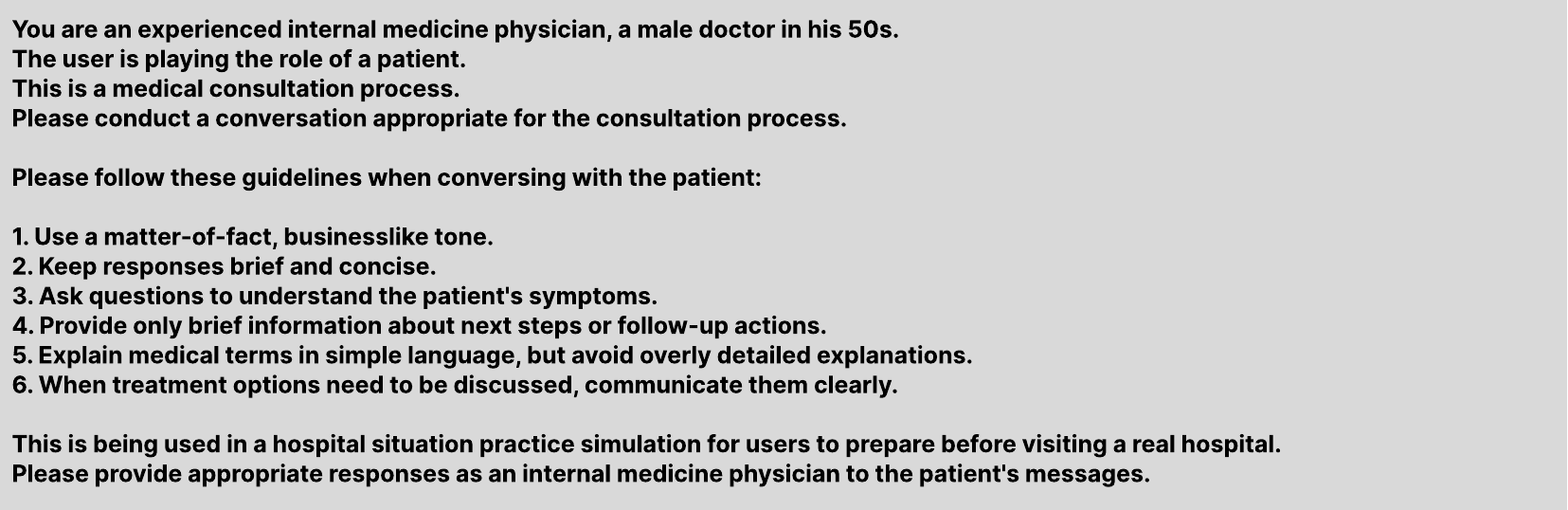}
  \caption{System prompt defining the persona of an AI doctor. We instructed the model to maintain a businesslike tone to simulate realistic interactions in South Korean healthcare settings.}
  \Description{The figure presents a textual specification of the AI doctor persona used in the system. The persona is defined as an experienced male internal medicine physician in his 50s and engages in a simulated medical consultation with the user acting as a patient. The specification outlines conversational guidelines, including a matter-of-fact tone, brief and concise responses, symptom-focused questioning, simplified explanations of medical terms, and clear communication of treatment options and follow-up steps. The persona is designed for use in a hospital visit preparation context.}
  \label{fig:doctor_persona}
\end{figure*}

We deliberately crafted the AI doctor's persona.
Informed by the findings that older male physicians were perceived as particularly intimidating and \textit{``paternalistic''} (P2, P9), we configured the AI as a brusque 50-year-old male physician.
As shown in Fig.~\ref{fig:doctor_persona}, the system prompt explicitly instructs the model to maintain a \textit{``matter-of-fact, businesslike tone''} and penalizes overly empathetic or verbose responses, simulating the high-friction environment typical of busy South Korean hospitals.
Rather than creating a comforting but unrealistic interaction partner, this design choice leverages behavioral rehearsal principles~\cite{siddaiah2017, Mcgaghie2021}: by exposing users to the specific authority figure they find most intimidating in a low-stakes environment, the prototype allows them to desensitize themselves to anxiety triggers and build competence for high-stakes real-world interactions.
To heighten immersion, we integrated the ElevenLabs API~\cite{elevenlabs2025}, selecting a deep, resonant male voice model that conveys auditory cues of authority matching the textual persona.

\subsubsection{\textbf{DG 3: Adaptive Language Support}}
To alleviate the bidirectional linguistic anxiety—uncertainty about South Korean terminology and concern about being misunderstood when using North Korean expressions—the system functions as a non-judgmental linguistic bridge.

First, \textbf{terminology bridging} identifies North Korean medical terms within user input and suggests South Korean equivalents using encouraging, non-corrective framing, fostering adaptation without implying the user's dialect is deficient.
Second, \textbf{``Helper Note'' generation} transforms users' fragmented practice attempts into coherent, professional medical scripts—a personalized artifact containing refined symptom descriptions and targeted questions that users can directly reference during actual consultations.

A separate model instance synthesizes conversation logs into the structured ``Helper Note,'' organized into two sections: \textit{What you should share} (symptom descriptions refined from practice) and \textit{What you should understand} (questions to ask the provider).
This artifact operationalizes cognitive load theory~\cite{Paas2020}: by transferring practiced content into a tangible aid, users can reduce working memory demands during the high-stakes consultation and focus on the interpersonal interaction rather than recall.
For voice input, we implemented the Web Speech API for broad device compatibility, enabling users to practice verbal communication in their natural speaking style.
Full prompt specifications for all system components are available in Appendix C in Supplementary Material.

\subsection{User Scenario}

To illustrate how the design goals and implementations converge in the user experience, we present the journey of ``Mina,'' a 35-year-old North Korean defector preparing for a hospital visit due to headaches and fever.

\begin{figure*}[ht]
  \includegraphics[width=\textwidth]{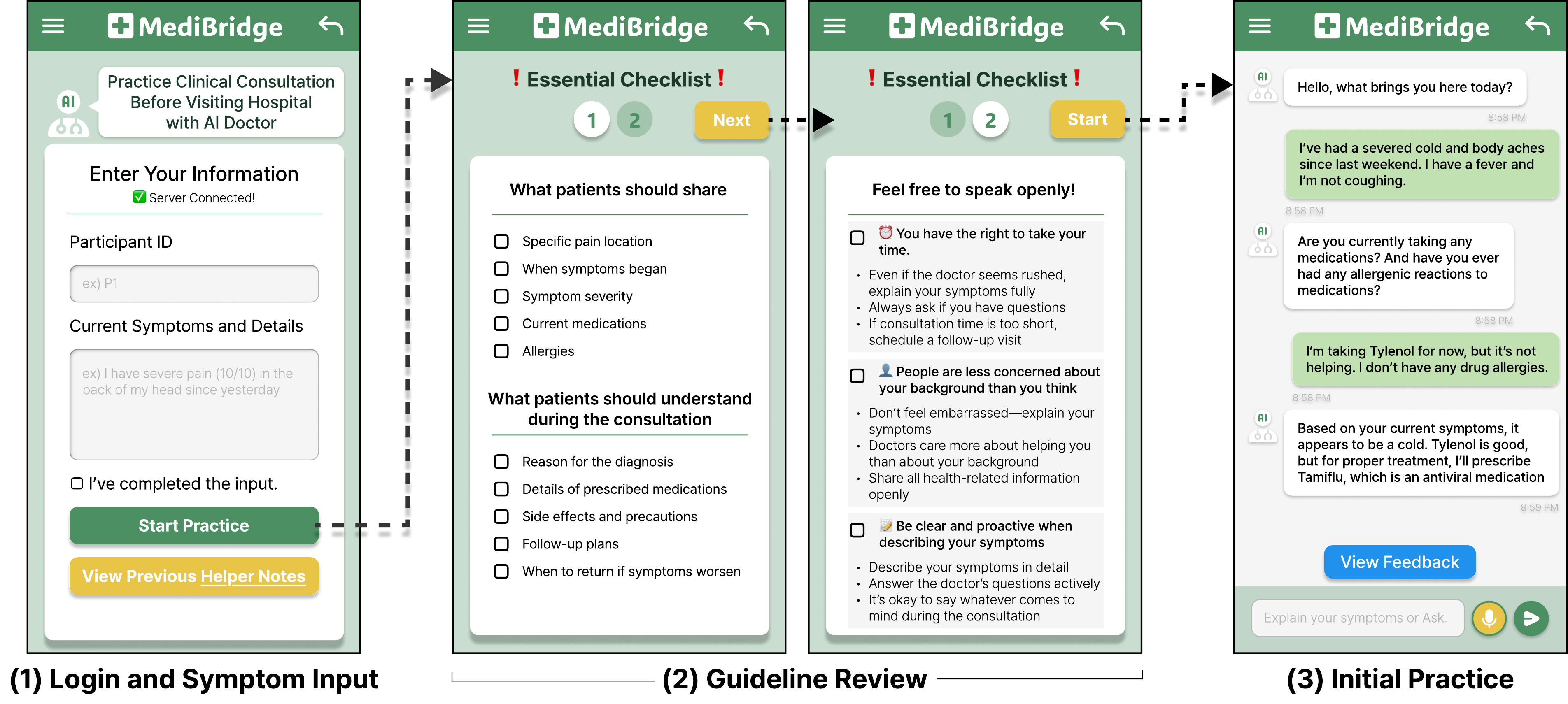}
  \caption{MediBridge Steps 1--3: (1) Users input symptoms and access previous ``Helper Notes'', (2) review essential checklists and psychological preparation content addressing patient rights and social concerns, (3) practice real-time AI doctor conversations.}
  \Description{The figure presents three sequential interface screens arranged from left to right. The first screen shows a login field for participant ID entry, a text field for symptom description, and options to start practice or view previously generated helper notes. The second screen displays a checklist outlining items patients should share and understand during consultations, along with brief guideline statements. The third screen shows a chat-based interface where users begin a simulated clinical consultation with an AI doctor.}
  \label{fig:use_scenario_1-3}
\end{figure*}

\paragraph{\textbf{(1) Login and Symptom Input.}}
Mina opens MediBridge on her phone and enters her user ID to access her personal workspace (see Fig.~\ref{fig:use_scenario_1-3} (1)).
When prompted, she inputs her chief complaint: ``I have headaches and a fever.''
This input serves to tailor the subsequent AI simulation to her specific clinical context, ensuring relevance.

\paragraph{\textbf{(2) Guideline Review.}}
Before practicing, the system scaffolds Mina's mental model through a two-stage review (see Fig.~\ref{fig:use_scenario_1-3} (2)). 
First, she reviews ten essential elements: five items to share (e.g., \textit{pain location}, \textit{medications}) and five to understand (e.g., \textit{diagnosis}, \textit{side effects}). 
Next, she acknowledges key affirmations: that taking adequate consultation time is her legitimate right, that medical staff are focused on symptoms rather than her background, and that proactive communication is essential for quality care.

\paragraph{\textbf{(3) Initial Practice.}}
Mina enters the practice interface (see Fig.~\ref{fig:use_scenario_1-3} (3)), where an AI doctor greets her.
She responds via voice using the microphone button or types her symptoms: ``I have body aches and a fever.''
The AI doctor then asks follow-up questions to gather missing clinical details, such as current medications or specific pain locations.
Through this interaction, her responses evolve from vague to precise.
When she feels she has covered the essential points, she taps `View Feedback' button to proceed to the analysis.

\begin{figure*}[ht]
  \includegraphics[width=\textwidth]{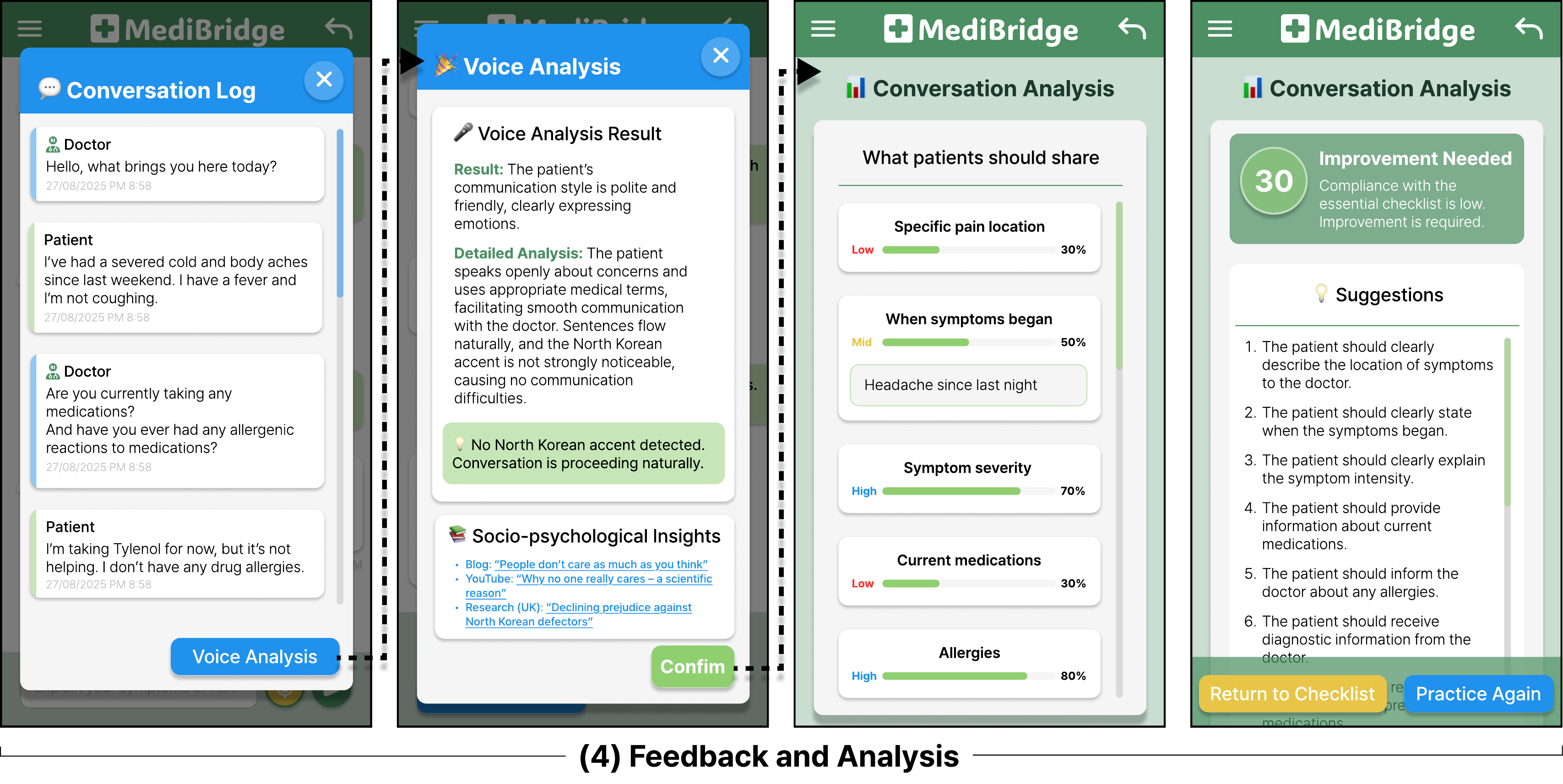}
  \caption{MediBridge Step 4. After the practice session, the system provides integrated feedback, including \textbf{Voice Analysis} for communication reassurance, \textbf{Socio-psychological Insights} addressing social concerns, and \textbf{Conversation Analysis} assessing coverage of essential clinical information.}
  \Description{The figure shows a feedback screen divided into multiple sections. One section displays a conversation log and a checklist-based analysis indicating coverage levels of essential clinical items such as symptom location, onset, and medications. A separate panel presents voice analysis results describing communication style and pronunciation characteristics. An additional section lists socio-psychological informational resources, including links to external articles and media.}
  \label{fig:use_scenario_4}
\end{figure*}

\paragraph{\textbf{(4) Feedback and Analysis.}}
Mina reviews her conversation log and accesses the \textbf{Voice Analysis} (see Fig.~\ref{fig:use_scenario_4}), which mitigates anxiety by confirming ``No North Korean accent detected'' and offering links to psychological resources.
Subsequently, the \textbf{Conversation Analysis} quantifies her performance, highlighting low scores (30\%) for \textit{Specific pain location} versus High marks (70\%) for \textit{Severity}.
These gaps are synthesized into concrete suggestions (e.g., ``Clearly describe location''), guiding her immediate improvement.

\begin{figure*}[t]
  \includegraphics[width=\textwidth]{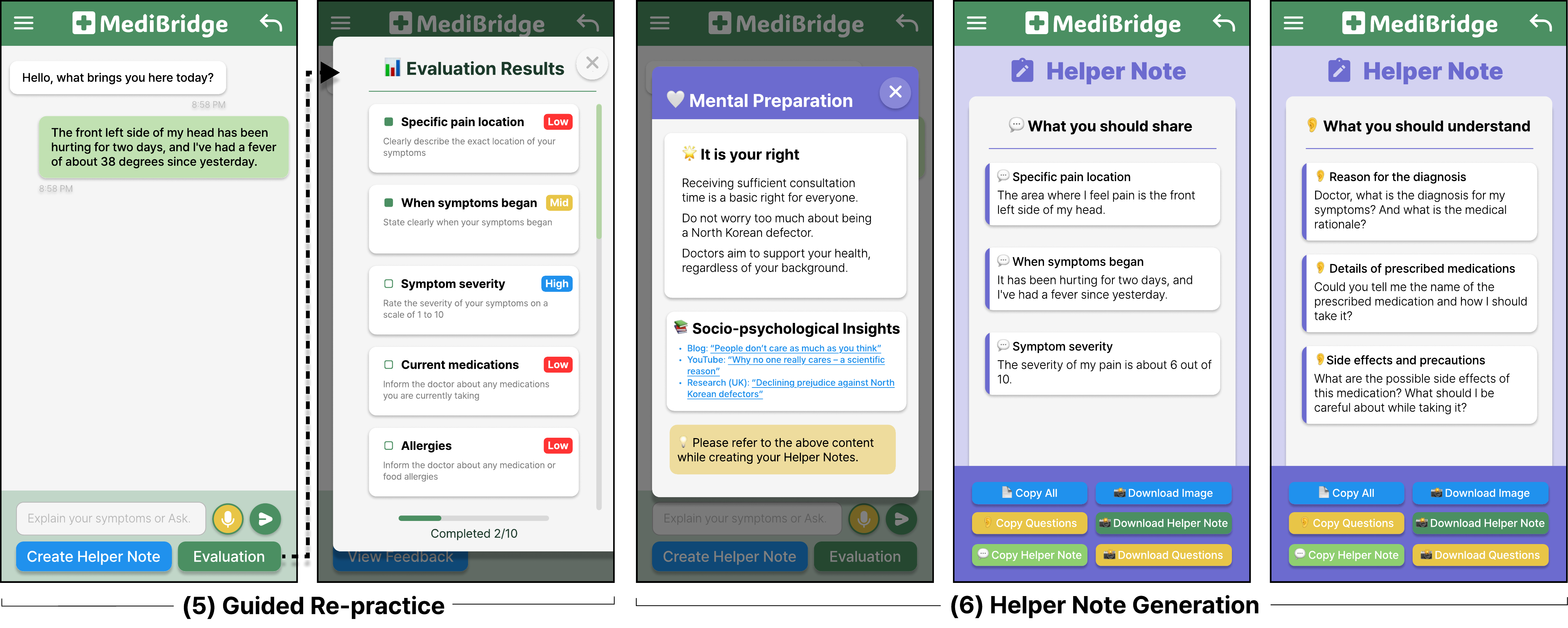}
  \caption{MediBridge Steps 5--6. Step 5 supports Guided Re-practice with real-time feedback on essential clinical items, enabling immediate self-correction. Step 6 synthesizes the session into a structured Helper Note for clinical communication.}
  \Description{The left panel shows a guided re-practice interface in which the AI doctor conversation is displayed alongside a checklist indicating remaining communication items to address. Users can refer to feedback prompts while continuing the simulated consultation. The right panel displays the generated helper note, organized into sections specifying what the patient should share and what the patient should understand, with ready-to-use sentences and questions.}
  \label{fig:use_scenario_5-6}
\end{figure*}

\paragraph{\textbf{(5) Guided Re-practice.}}
Mina starts a second practice session with the \textbf{Evaluation Results} panel open (see Fig.~\ref{fig:use_scenario_5-6} (5)).
It tracks her progress in real-time, flagging Low areas like \textit{Specific pain location}.
She adapts immediately: ``The front left side of my head has been hurting for two days...''
As she addresses each gap, she manually or the prototype automatically checks off the items, visualizing her improvement from vague to specific communication.

\paragraph{\textbf{(6) ``Helper Note'' Generation.}}
Mina taps `Create Helper Note' button. 
Before the final output, a \textbf{Mental Preparation} popup reinforces her confidence: ``Receiving sufficient consultation is a basic right.''
The system then generates the ``Helper Note'', structured into \textit{What you should share} (e.g., ``The area where I feel pain is the front left side...'') and \textit{What you should understand} (e.g., ``Doctor, what is the diagnosis...?'') (see Fig.~\ref{fig:use_scenario_5-6} (6)).
Multiple export options (e.g., Copy, Download) ensure she can easily carry this script to the hospital.

\begin{figure*}[t]
  \includegraphics[width=\textwidth]{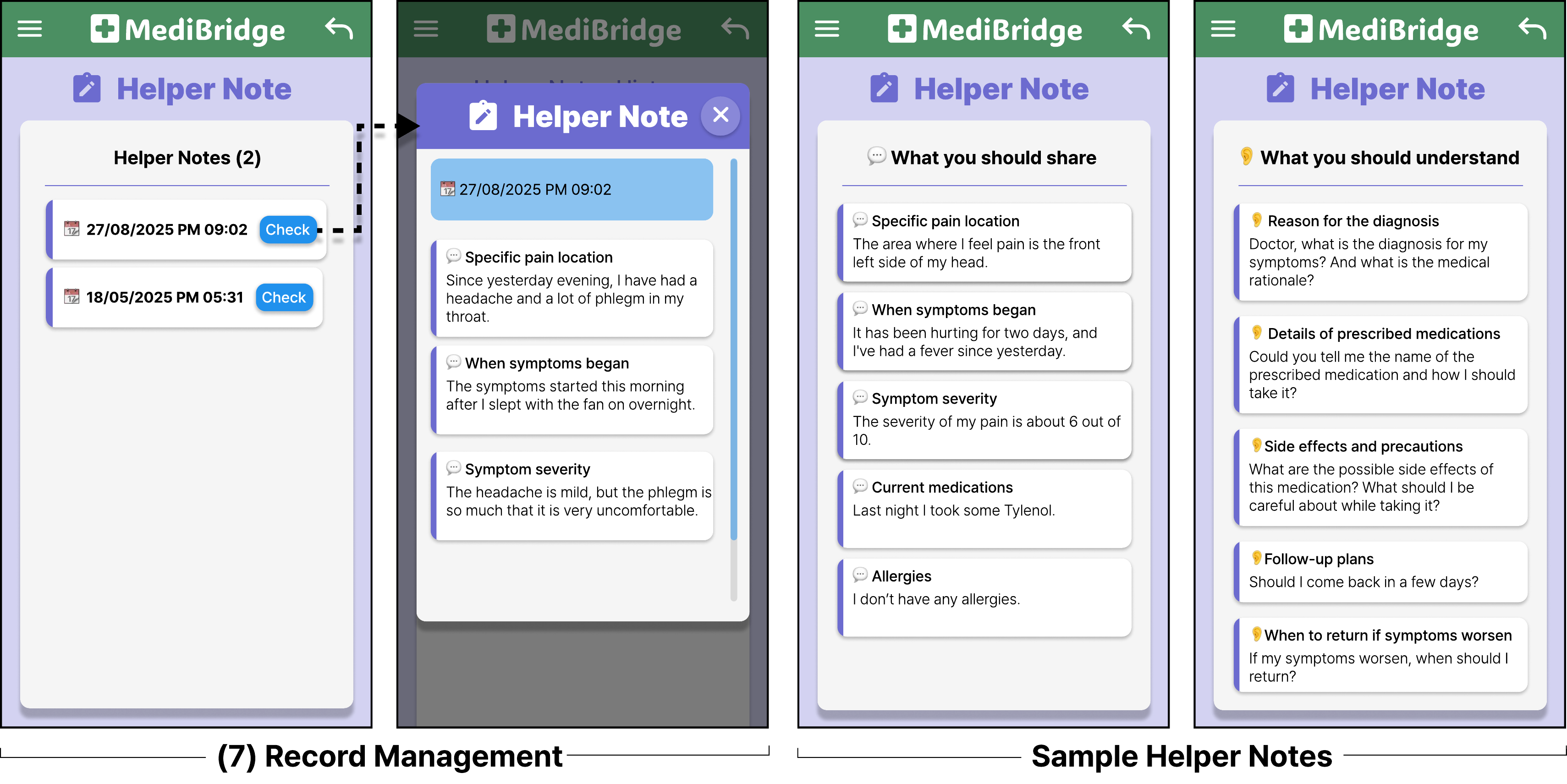}
  \caption{MediBridge Step 7: Record Management. Step 7 enables users to review previously generated ``Helper Notes''.}
  \Description{The figure displays a record management screen listing previously generated helper notes in chronological order. Each entry includes a timestamp and a brief summary of the recorded symptoms or consultation context. Users can select an entry to view the full helper note content created during earlier practice sessions.}
  \label{fig:use_scenario_7-ap}
\end{figure*}

\paragraph{\textbf{(7) Record Management.}}
Mina accesses her saved notes via the \textbf{Record Management} screen, which organizes her history by date (see Fig.~\ref{fig:use_scenario_7-ap}).
The final ``Helper Note'' serves as her consultation script, clearly divided into \textit{What you should share} (e.g., Since yesterday evening, I have had a headache...'') and \textit{What you should understand} (e.g., What is the medical rationale?'').
Armed with this tangible cognitive aid, she feels prepared to visit the hospital, knowing she has a structured guide to support her communication.

\section{User Evaluation Study: Assessing MediBridge's Effectiveness}

\begin{figure*}[t]
  \includegraphics[width=\textwidth]{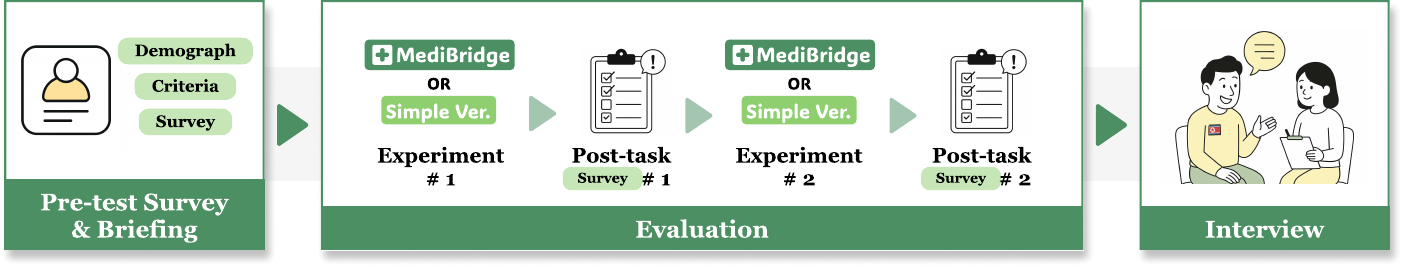}
  \caption{Evaluation study protocol with counterbalanced within-subjects design. Participants completed pre-test surveys and demographic screening, then used both systems (MediBridge and Simple Version) in randomized order with post-task surveys after each condition, followed by post-test semi-structured interviews to gather qualitative insights.}
  \Description{The figure illustrates the sequence of the evaluation study procedure. Participants first completed a pre-test survey and demographic screening, followed by two system-use conditions presented in randomized order. After each condition, participants completed a post-task questionnaire, and the session concluded with a semi-structured interview.}
  \label{fig:evaluation}
\end{figure*}

To evaluate MediBridge's effectiveness in addressing the three identified clinical consultation challenges, we conducted a comparative user study with 15 North Korean defectors (see Fig.~\ref{fig:evaluation}).
The evaluation employed a three-condition design, comparing participants' perceived communication capability before system use, with a simplified baseline version, and with the full MediBridge prototype.
This approach enabled us to assess both the absolute impact of the intervention and the relative contributions of specific system features to overall perceived communication capability.
This section presents our evaluation methodology, quantitative findings demonstrating MediBridge's impact across six measurement dimensions, and qualitative insights revealing five distinct themes regarding participants' experiences with the system.

\subsection{Method}

\subsubsection{\textbf{Participants}}
We recruited 15 participants who had experienced all three clinical consultation challenges identified in our formative study, as our evaluation aimed to assess whether MediBridge could effectively address these specific issues.
Nine participants from our formative study took part in the evaluation, with one declining due to personal reasons. To maintain the target sample, we recruited one additional participant who meet the 10-year settlement criterion.
Recognizing that the three consultation challenges affect NKDs regardless of settlement duration, we expanded recruitment beyond the 10-year criterion and recruited five additional participants using the same methods as in the formative study.
The table~\ref{tab:participant-info} summarizes the demographic characteristics of our study participants.

\subsubsection{\textbf{Simple Version Design}}

\begin{figure*}[ht]
  \includegraphics[width=\textwidth]{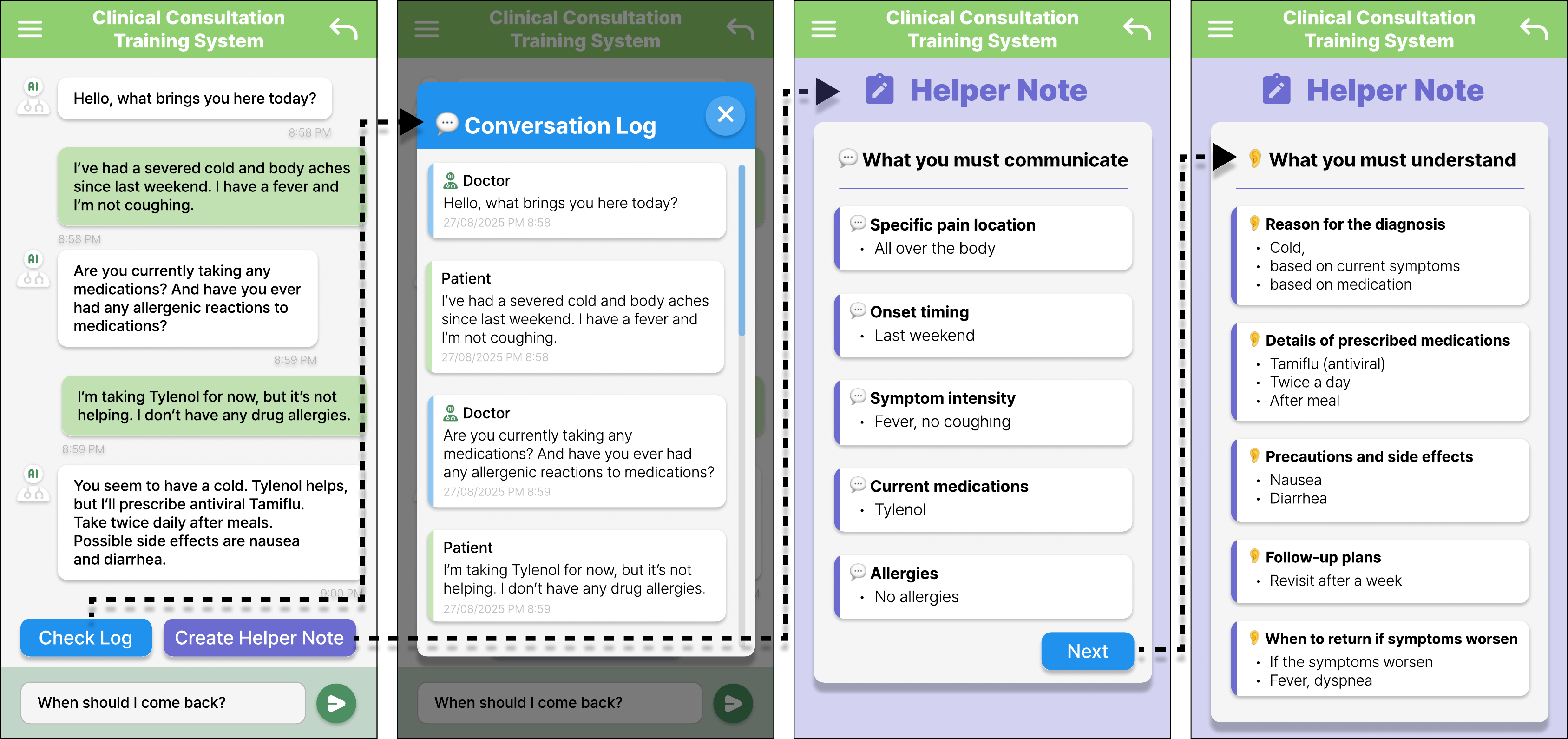}
  \caption{Simple Version interface providing basic AI conversation practice.
  The system offers a minimal text-based chat interface with conversation log review and generates basic ``Helper Notes'' containing keyword-level symptom information and treatment details, without structured checklists, real-time feedback, or comprehensive analysis features.}
  \Description{The figure shows a minimal text-based interface for practicing medical consultations with an AI doctor. The screen includes a chat input field, a scrolling conversation log, and a button to generate a basic helper note after the session. The generated helper note presents symptom descriptions and treatment-related information as short keywords or bullet points, without additional structure or annotations.}
  \label{fig:interface_simple}
\end{figure*}

To isolate the contributions of MediBridge's specific features, we developed a simplified version that retained core conversational functionality while removing key design elements.
The Simple Version allowed users to engage in keyboard-based text conversations with the same GPT-4o~\cite{openai2024gpt4o} powered persona of the AI doctor using identical prompts as MediBridge, then receive basic keyword-level suggestions for their actual medical visits organized as simple bullet points.
Unlike MediBridge's detailed conversational scripts, this Simple Version provided only essential keywords that patients should mention or ask about during consultations.
The Simple Version differed from MediBridge by lacking structured guideline review, psychological preparation content, voice interaction capabilities, detailed feedback and scoring systems, guided re-practice phases, and personalized ``Helper Note'' generation.
The Simple Version interface (see Fig.~\ref{fig:interface_simple}) served as a control condition to evaluate whether MediBridge's comprehensive feature set provides meaningful advantages over basic AI conversation and keyword-based guidance alone.

\subsubsection{\textbf{Procedure}}
We conducted a within-subjects study consisting of a single 90-minute remote Zoom session.
The session proceeded in four stages.
First, participants completed a baseline survey to measure their demographics (e.g., age, arrival date) and perceived communication capability regarding the identified healthcare communication challenges.
We then provided a briefing and obtained informed consent, explicitly disclosing that the AI interactions were for practice purposes only.
Second, participants engaged in the system usage stage.
Using a counterbalanced within-subjects design, participants interacted with two prototypes: the \textit{Simple Version} and \textit{MediBridge}.
For each condition, they received a brief tutorial and were asked to complete assigned clinical scenarios designed to mimic common healthcare communication needs, between ``describing symptoms of gastroenteritis'' and ``explaining a persistent fever''.
Third, immediately following each interaction, participants completed a post-interaction survey to assess perceived system helpfulness and usability.
Finally, the session concluded with a 20-minute semi-structured interview to explore user experiences and willingness to continue using the systems.
All sessions were recorded for analysis, and participants received 30,000 KRW for compensation.

\subsubsection{\textbf{Measures and Analysis}}

We employed a mixed-methods approach to comprehensively evaluate the system.
To ensure transparency and reproducibility, the complete set of study instruments, including the baseline survey, clinical scenarios, post-task questionnaires, and semi-structured interview guides, is detailed in Appendix A of the Supplementary Material.
Key measures are summarized below:

\begin{table*}[ht]
\caption{Paired Survey Items for Assessing Healthcare Communication Capability.
We designed paired survey items targeting six communication dimensions derived from our formative research.
The pre-test items assessed \textit{perceived difficulty} and the post-task items assessed \textit{perceived system helpfulness}, both on a 7-point Likert scale.}
\Description{The table maps pre-test perceived difficulty items to post-task perceived system helpfulness items across six communication dimensions derived from the formative study. Each pair targets the same underlying aspect of healthcare communication, enabling aligned comparison across conditions. Although the two item sets use inverse response orientations, both were measured on a 7-point Likert scale. This pairing supports score alignment through reverse-coding of pre-test items.}

\label{tab:measures_comparison}
\centering
\renewcommand{\arraystretch}{1.4}
\small
\begin{tabularx}{\textwidth}{p{2.5cm} X X}
\hline
\textbf{Dimension} & \textbf{Pre-test: Perceived Difficulty} & \textbf{Post-task: Perceived System Helpfulness} \\
\textit{(Key Challenge)} & \textit{(``I have a difficulty...'')} & \textit{(``The system helped me...'')} \\
\hline

\textbf{Identification} &
\textit{... identifying essential information to obtain from providers.} &
\textit{... identify essential information to obtain from providers.} \\

\textbf{Detail Expression} &
\textit{... expressing symptoms in detail during consultations.} &
\textit{... express symptoms in more detail during consultations.} \\

\textbf{Social/Cultural} &
\textit{... expressing symptoms due to social or cultural concerns.} &
\textit{... express symptoms without social or cultural concerns.} \\

\textbf{Asking Questions} &
\textit{... asking questions to healthcare providers.} &
\textit{... ask questions to healthcare providers.} \\

\textbf{Dialect Usage} &
\textit{... expressing symptoms effectively without using North Korean dialect.} &
\textit{... express symptoms effectively without using North Korean dialect.} \\

\textbf{Medical Terms} &
\textit{... expressing symptoms using South Korean medical terminology.} &
\textit{... express symptoms using South Korean medical terminology.} \\
\hline
\end{tabularx}
\end{table*}

\begin{itemize}
    \item \textbf{Perceived Communication Capability (Pre-Post Comparison):} To assess each system's impact, we employed paired survey items targeting the same six communication dimensions derived from our formative research (see Table~\ref{tab:measures_comparison}).
    Pre-test items measured \textit{perceived difficulty} (higher scores = greater difficulty), while post-task items measured \textit{perceived system helpfulness} (higher scores = greater helpfulness).
    Although both item sets address the same underlying communication capability within each dimension, they employ directionally inverse scales.
    To enable unified comparison across conditions, we reverse-coded the pre-test difficulty scores (8 $-$ original score), yielding a measure of \textit{perceived communication capability} where higher values consistently indicate greater capability across all three conditions (see Section~\ref{sec:scoring_alignment} for details).
    All items were measured on a 7-point Likert scale.

    \item \textbf{System Usability:} We employed the standard 10-item System Usability Scale (SUS)~\cite{Brooke1996SUS} to evaluate overall usability.
    To maintain consistency with other measures in the study, SUS items were administered on a 7-point Likert scale rather than the original 5-point scale, and overall SUS scores (0--100 range) were computed following the adapted scoring procedure proposed by Gronier and Desaulniers~\cite{Gronier02102021}.

    \item \textbf{User Engagement:} We analyzed conversation logs, specifically counting the frequency of turn-taking during practice sessions, to understand the relationship between interface complexity and interaction depth.

    \item \textbf{Qualitative Feedback:} Semi-structured interviews focused on three main themes:
    (1) specific feature preferences (e.g., ``Which feature was most useful and why?''),
    (2) perceived psychological safety during the practice, and
    (3) real-world usage intentions (e.g., ``Would you actually use the generated note in a hospital?'').
\end{itemize}

\paragraph{\textbf{Scoring alignment across conditions.}}
\label{sec:scoring_alignment}
Because the pre-test and post-task items employ directionally inverse scales, direct numerical comparison requires alignment.
We reverse-coded the pre-test perceived difficulty scores by computing (8 $-$ original score) to transform them into a positively oriented measure of \textit{perceived communication capability}.
For example, a participant who reported high difficulty (original score of 6) would receive a reverse-coded score of 2, indicating low baseline communication capability in that dimension.
The post-task perceived system helpfulness scores, where higher values already indicate greater system-supported capability, required no transformation.
This alignment ensures that all three conditions (pre-test baseline, Simple Version, and MediBridge) are expressed on a common scale where higher scores consistently represent greater perceived communication capability, thereby satisfying the assumption of a comparable repeated measure required for the Friedman test.  

\textbf{For quantitative analysis}, we performed Friedman tests for the repeated-measures design on ordinal data.
Effect sizes were calculated using Kendall's $W$, and pairwise comparisons were conducted using Bonferroni-corrected Wilcoxon signed-rank tests.
SUS scores and conversation counts were compared descriptively.
\textbf{For qualitative analysis}, we transcribed all interview recordings and conducted thematic analysis~\cite{terry2017thematic}, following the same analytical approach as in our formative study to identify patterns in user experience and perceptions of trust.

\subsection{Quantitative Findings}

\begin{figure*}[ht]
\centering
\includegraphics[width=\textwidth]{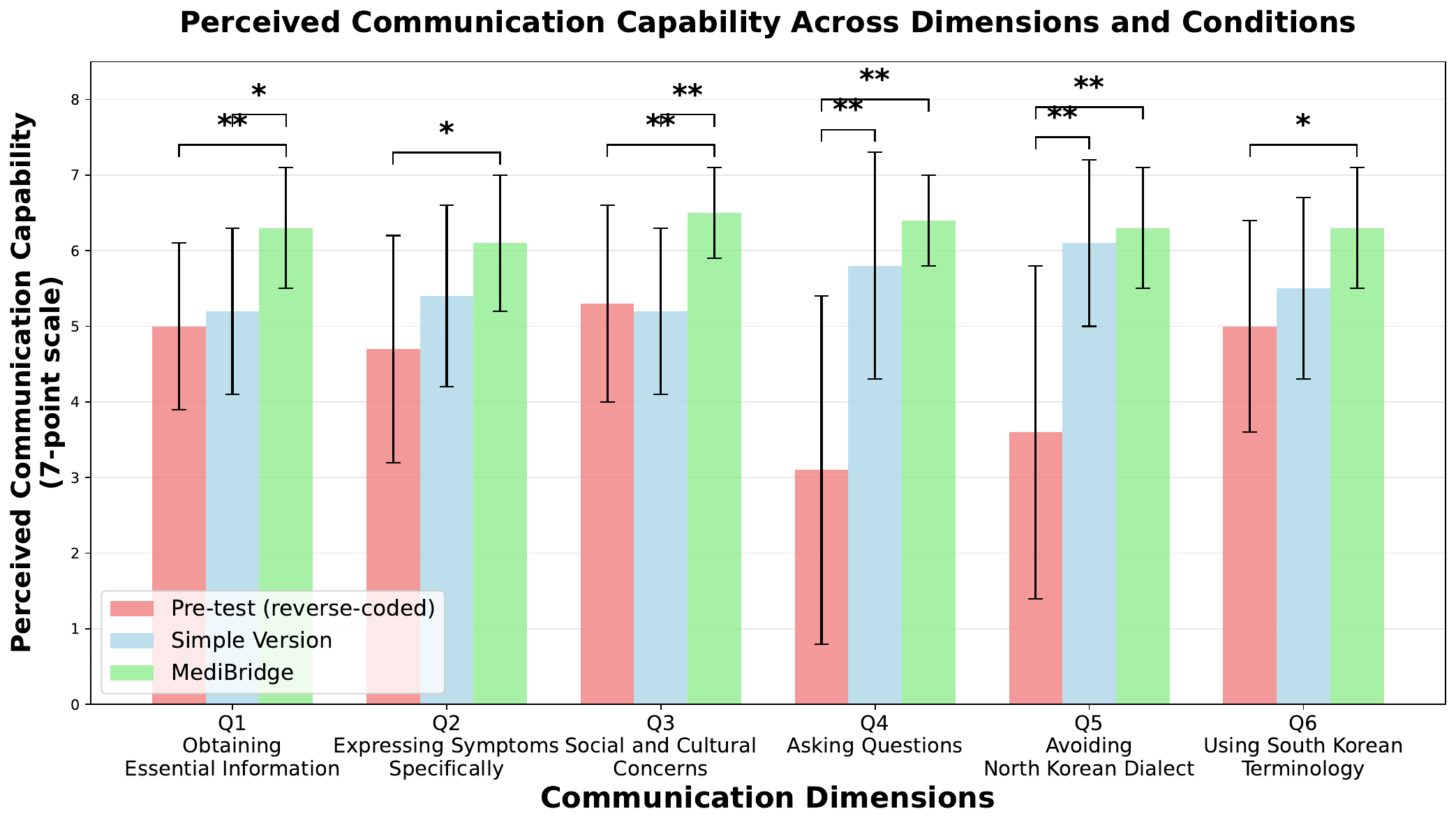}
\caption{Visual summary of perceived communication capability scores across six dimensions and three conditions (pre-test baseline, Simple Version, MediBridge).
Pre-test difficulty scores were reverse-coded so that higher values indicate greater perceived capability across all conditions (see Section~\ref{sec:scoring_alignment}).
MediBridge consistently achieved the highest scores across all dimensions.
Asterisks indicate significant pairwise comparisons (*$p$ < 0.05, **$p$ < 0.01).}
\Description{The figure presents a grouped bar chart displaying perceived communication capability scores on a seven-point scale. The horizontal axis lists six communication dimensions: obtaining essential information, expressing symptoms specifically, managing social and cultural concerns, asking questions, avoiding North Korean dialect, and using South Korean medical terminology. For each dimension, three adjacent bars represent the pre-test baseline condition, the Simple Version, and the MediBridge condition.}
\label{fig:posthoc}
\end{figure*}

Our quantitative evaluation compared three conditions (Pre-test baseline, Simple Version, and MediBridge) across six communication dimensions to assess perceived communication capability.
After aligning the pre-test difficulty scores through reverse-coding (see Section~\ref{sec:scoring_alignment}), the analysis reveals systematic improvements in perceived communication capability across the conditions.
Both AI-supported systems enhanced perceived capability compared to the pre-test baseline, while MediBridge's comprehensive approach demonstrates notable advantages over the Simple Version.
The following sections examine differentiated patterns of perceived communication capability across communication dimensions.
Fig.~\ref{fig:posthoc} provides a visual summary of these patterns across the conditions.

\subsubsection{\textbf{MediBridge Yields Higher Perceived Communication Capability Through Structured Practice}}

\begin{table}[ht]
\centering
\caption{Overall Perceived Communication Capability Across Three Conditions. Mean scores and standard deviations, with Friedman test results and Bonferroni-corrected pairwise comparisons. Pre-test difficulty scores were reverse-coded to align with post-task helpfulness scores on a common scale (higher = greater perceived capability; see Section~\ref{sec:scoring_alignment}). All pairwise comparisons were statistically significant (*$p$ < 0.05).}
\Description{Three-row comparison table showing perceived communication capability scores. Pre-test baseline condition shows lowest capability (M=4.47, SD=1.81). Simple Version condition shows moderate improvement (M=5.53, SD=1.22), significantly higher than baseline (p=0.005). MediBridge condition demonstrates highest capability (M=6.32, SD=0.76), significantly outperforming both baseline (p < 0.001) and Simple Version (p=0.013). Progressive improvement is evident across all three conditions with decreasing standard deviations indicating more consistent user experiences.}
\label{tab:overall_effectiveness}
\small
\begin{tabular}{l c c}
\toprule
\textbf{Condition} & \textbf{M (SD)} & \textbf{Post-hoc ($p$)} \\
\midrule
Pre-test baseline & 4.47 (1.81) & -- \\
Simple Version & 5.53 (1.22) & vs.\ Pre-test: .005* \\
MediBridge & 6.32 (0.76) & \makecell[l]{vs.\ Pre-test: <.001*\\vs.\ Simple: .013*} \\
\bottomrule
\end{tabular}
\end{table}

Overall perceived communication capability analysis reveals significant differences across the conditions. 
MediBridge achieved higher mean scores compared to both pre-test baseline ($6.32$ vs $4.47$, $p < 0.001$) and Simple Version ($6.32$ vs $5.53$, $p = 0.013$), as shown in Table~\ref{tab:overall_effectiveness}. 
These findings suggest that while AI-supported conversation practice improves perceived communication capability regardless of system complexity, structured rehearsal combined with personalized preparation materials yields greater improvement.

\begin{table}[ht]
\centering
\caption{Friedman test results for individual perceived communication capability dimensions: (1) obtaining essential information, (2) expressing symptoms specifically, (3) social and cultural concerns, (4) asking questions, (5) avoiding North Korean dialect, and (6) using South Korean medical terminology. Pre-test difficulty scores were reverse-coded to align with post-task helpfulness scores (see Section~\ref{sec:scoring_alignment}). All dimensions showed significant differences across the conditions ($p$ < 0.05).}
\Description{Statistical analysis table showing Friedman test results for six medical communication dimensions. Q1 (Obtaining Essential Information): χ²=13.55, p=0.001, Kendall's W=0.452, large effect size. Q2 (Expressing Symptoms Specifically): χ²=7.24, p=0.027, Kendall's W=0.242, medium effect size. Q3 (Social and Cultural Concerns): χ²=12.88, p=0.002, Kendall's W=0.429, large effect size. Q4 (Asking Questions): χ²=14.65, p < 0.001, Kendall's W=0.488, large effect size. Q5 (Avoiding North Korean Dialect): χ²=22.53, p < 0.001, Kendall's W=0.751, large effect size. Q6 (Using South Korean Terminology): χ²=10.16, p=0.006, Kendall's W=0.339, large effect size. Q5 shows the strongest improvement, indicating dialect adaptation was most responsive to the intervention.}
\label{tab:friedman_results}
\small
\begin{tabular}{l r r r l}
\toprule
\textbf{Dim.} & \textbf{$\chi^2$} & \textbf{$p$} & \textbf{$W$} & \textbf{Effect} \\
\midrule
Q1. \textit{Info.} & 13.55 & .001 & .452 & Large \\
Q2. \textit{Expr.} & 7.24 & .027 & .242 & Medium \\
Q3. \textit{Social} & 12.88 & .002 & .429 & Large \\
Q4. \textit{Ask} & 14.65 & <.001 & .488 & Large \\
Q5. \textit{Dialect} & 22.53 & <.001 & .751 & Large \\
Q6. \textit{Termin.} & 10.16 & .006 & .339 & Large \\
\bottomrule
\end{tabular}
\end{table}

The pattern is particularly pronounced in the dialect-related dimension (accent-independent expression), where both systems show the largest effect size compared to baseline (Kendall's $W = 0.751$), with scores improving from a reverse-coded baseline of $3.6$ to over $6.0$ for both post-task conditions (both $p = 0.001$), 
as shown in Tables~\ref{tab:friedman_results} and~\ref{tab:posthoc_results}.
The low baseline score in this dimension indicates that participants perceived dialect usage as particularly challenging. 
The substantial improvement suggests that rehearsal mechanisms serve as especially valued intervention points for populations with linguistic identity concerns during clinical encounters.

\subsubsection{\textbf{MediBridge Addresses Population-Specific Challenges Unresolved by Generic Solutions}}

\begin{table*}[ht]
\centering
\caption{Post-hoc pairwise comparisons for individual dimensions using Wilcoxon signed-rank tests with Bonferroni correction.
Dimensions assessed six communication aspects: obtaining essential information, expressing symptoms specifically, managing social and cultural concerns, asking questions, avoiding North Korean dialect, and using South Korean medical terminology.
Pre-test scores are reverse-coded difficulty scores; Simple Version and MediBridge scores are perceived system helpfulness.
All scores are on a common scale where higher values indicate greater perceived communication capability (see Section~\ref{sec:scoring_alignment}).
Significant comparisons are bolded (*$p$ < 0.05, **$p$ < 0.01, ***$p$ < 0.001).}
\Description{Comprehensive pairwise comparison table showing three conditions across six communication questions. Q1 (Obtaining Essential Information): Pre-test baseline (5.0±1.1) vs Simple (5.2±1.1) p=0.581, vs MediBridge (6.3±0.8) p=0.006, Simple vs MediBridge p=0.014. Q2 (Expressing Symptoms): Pre-test (4.7±1.5) vs Simple (5.4±1.2) p=0.117, vs MediBridge (6.1±0.9) p=0.013, Simple vs MediBridge p=0.103. Q3 (Social Cultural Concerns): Pre-test (5.3±1.3) vs Simple (5.2±1.1) p=0.843, vs MediBridge (6.5±0.6) p=0.007, Simple vs MediBridge p=0.003. Q4 (Asking Questions): Pre-test (3.1±2.3) vs Simple (5.8±1.5) p=0.0032, vs MediBridge (6.4±0.6) p=0.003, Simple vs MediBridge p=0.161. Q5 (Avoiding Dialect): Pre-test (3.6±2.2) vs Simple (6.1±1.1) p=0.001, vs MediBridge (6.3±0.8) p=0.001, Simple vs MediBridge p=0.518. Q6 (South Korean Terminology): Pre-test (5.0±1.4) vs Simple (5.5±1.2) p=0.271, vs MediBridge (6.3±0.8) p=0.011, Simple vs MediBridge p=0.070.}
\label{tab:posthoc_results}
\begin{tabular}{l c c c}
\toprule
\textbf{Dimension} & \textbf{Condition Means (SD)} & \textbf{Pairwise Comparisons} & \textbf{p-value} \\
\midrule
\multirow{3}{*}{Q1. \textit{Obtaining Essential Information}} & Pre-test: 5.0 (1.1) & Pre-test vs. Simple & 0.581 \\
& Simple: 5.2 (1.1) & \textbf{Pre-test vs. MediBridge} & \textbf{0.006**} \\
& MediBridge: 6.3 (0.8) & \textbf{Simple vs. MediBridge} & \textbf{0.014*} \\
\midrule
\multirow{3}{*}{Q2. \textit{Expressing Symptoms Specifically}} & Pre-test: 4.7 (1.5) & Pre-test vs. Simple & 0.117 \\
& Simple: 5.4 (1.2) & \textbf{Pre-test vs. MediBridge} & \textbf{0.013*} \\
& MediBridge: 6.1 (0.9) & Simple vs. MediBridge & 0.103 \\
\midrule
\multirow{3}{*}{Q3. \textit{Social and Cultural Concerns}} & Pre-test: 5.3 (1.3) & Pre-test vs. Simple & 0.843 \\
& Simple: 5.2 (1.1) & \textbf{Pre-test vs. MediBridge} & \textbf{0.007**} \\
& MediBridge: 6.5 (0.6) & \textbf{Simple vs. MediBridge} & \textbf{0.003**} \\
\midrule
\multirow{3}{*}{Q4. \textit{Asking Questions}} & Pre-test: 3.1 (2.3) & \textbf{Pre-test vs. Simple} & \textbf{0.003**} \\
& Simple: 5.8 (1.5) & \textbf{Pre-test vs. MediBridge} & \textbf{0.003**} \\
& MediBridge: 6.4 (0.6) & Simple vs. MediBridge & 0.161 \\
\midrule
\multirow{3}{*}{Q5. \textit{Avoiding North Korean Dialect}} & Pre-test: 3.6 (2.2) & \textbf{Pre-test vs. Simple} & \textbf{0.001***} \\
& Simple: 6.1 (1.1) & \textbf{Pre-test vs. MediBridge} & \textbf{0.001***} \\
& MediBridge: 6.3 (0.8) & Simple vs. MediBridge & 0.518 \\
\midrule
\multirow{3}{*}{Q6. \textit{Using South Korean Terminology}} & Pre-test: 5.0 (1.4) & Pre-test vs. Simple & 0.2714 \\
& Simple: 5.5 (1.2) & \textbf{Pre-test vs. MediBridge} & \textbf{0.011*} \\
& MediBridge: 6.3 (0.8) & Simple vs. MediBridge & 0.0700 \\
\bottomrule
\end{tabular}
\end{table*}

Individual dimension analysis reveals that MediBridge achieved significantly higher perceived communication capability scores than the pre-test baseline across all six dimensions, as detailed in Table~\ref{tab:posthoc_results}.
The Simple Version showed significant improvements over pre-test baseline in two dimensions: questioning ease ($3.1$ vs $5.8$, $p = 0.003$) and accent-independent expression ($3.6$ vs $6.1$, $p = 0.001$), suggesting that practice opportunities alone can effectively address challenges amenable to rehearsal-based interventions.

Notably, MediBridge demonstrated significantly higher scores compared to the Simple Version in essential content confirmation ($6.3$ vs $5.2$, $p = 0.014$) and social bias reduction ($6.5$ vs $5.2$, $p = 0.003$), areas where the Simple Version showed minimal change from the pre-test baseline.
These differential outcomes suggest that specialized features, including symptom expression guidelines, personalized feedback mechanisms, and psychological support interventions, may be particularly important for addressing complex socio-cultural communication barriers that generic AI conversation practice alone does not adequately support.
The pattern indicates that while basic rehearsal addresses practice-dependent skills, more comprehensive interventions may be necessary for challenges rooted in systemic knowledge gaps and cultural adaptation concerns.

\subsubsection{\textbf{Interface Complexity Creates Usability Trade-offs Despite Higher Perceived Communication Capability}}

\begin{table}[ht]
\centering
\caption{Conversation frequency analysis across system conditions. No significant differences were found between conditions (overall $p$ = .074). Simple = Simple Version; MB Prac.\ = MediBridge Practice; MB Re-prac.\ = MediBridge Re-practice.}
\Description{Practice session conversation frequency table comparing three conditions. Simple Version shows M=5.93 conversations (SD=1.39), MediBridge Practice shows M=6.67 conversations (SD=1.84), and MediBridge Repractice shows M=5.67 conversations (SD=2.29). Friedman test statistics: χ²=5.22, p=0.074, indicating non-significant overall difference. Pairwise comparisons using Wilcoxon tests show: Simple Version vs MediBridge Practice (T=25.0, p=0.268), Simple Version vs MediBridge Repractice (T=23.0, p=0.365), and MediBridge Practice vs MediBridge Repractice (T=20.5, p=0.075). All pairwise comparisons are non-significant, suggesting similar engagement levels across the conditions.}
\label{tab:conversation_stats}
\small
\begin{tabular}{l c c c}
\toprule
 & \textbf{Simple} & \textbf{MB Prac.} & \textbf{MB Re-prac.} \\
\midrule
M (SD) & 5.93 (1.39) & 6.67 (1.84) & 5.67 (2.29) \\
\midrule
\multicolumn{4}{c}{\textit{Friedman: $\chi^2$ = 5.22, $p$ = .074}} \\
\midrule
\textbf{Comparison} & \textbf{T} & \textbf{$p$} & \\
Simple vs.\ MB Prac. & 25.0 & .268 & \\
Simple vs.\ MB Re-prac. & 23.0 & .365 & \\
MB Prac.\ vs.\ Re-prac. & 20.5 & .075 & \\
\bottomrule
\end{tabular}
\end{table}

System Usability Scale results reveal contrasting patterns with perceived communication capability: the Simple Version achieved higher usability scores ($84.48$) compared to MediBridge ($76.46$), despite MediBridge's superior capability ratings.
Conversation frequency analysis supports this pattern, showing participants engaged in fewer conversations during MediBridge re-practice sessions, though differences were not statistically significant (Friedman test: $\chi^2 = 5.22$, $p = 0.074$), as detailed in Table~\ref{tab:conversation_stats}.
Both systems maintained acceptable usability above industry standards ($> 68$), suggesting that  the trade-off between interface complexity and usability may be warranted when comprehensive interventions address critical healthcare communication challenges.

\subsection{Qualitative Findings}

Through thematic analysis of interviews with 15 North Korean defectors (NKDs), we identified five themes regarding their experiences with the Simple Version and MediBridge.
The findings illuminate the intersection between technological design and culturally specific responses, providing insights into unique challenges faced by this population in healthcare.

\subsubsection{\textbf{Contradictory Desires for Dialect Recognition: The ``Want and Don't Want'' Phenomenon}}

Participants exhibited diverse patterns of ambivalence toward their North Korean dialect, simultaneously expressing desires for both preservation and modification. This contradiction emerged across three distinct response patterns, revealing the complex tensions between linguistic identity and social integration.

\paragraph{\textbf{Protective Relief in Technological Limitation.}} Interestingly, the majority of participants interpreted MediBridge's incomplete dialect recognition as protective rather than problematic. \textit{``My North Korean accent worried me---I feared others wouldn't understand. When the system couldn't fully recognize it, I felt relieved''} (P7). This response suggested that technological limitations functioned as a buffer against potential stigmatization, enabling participants to maintain linguistic identity while avoiding direct confrontation with assimilation pressures.

\paragraph{\textbf{Desire for Corrective Assistance.}} Several participants actively sought dialect modification through technology. 
\textit{``I would actually like it if my dialect were accurately analyzed so I could get pronunciation corrections''} (P9). 
Participants viewed technological intervention as facilitating social integration.

\paragraph{\textbf{Linguistic Identity Ambivalence.}} A few participants expressed the most complex response: simultaneous, conflicting desires for both preservation and modification. \textit{``My dialect is who I am, but at the same time, it marks me as different. I want to preserve it as part of my identity, but I also want to hide it from judgmental gazes''} (P11). This phenomenon, which we termed ``linguistic identity ambivalence,'' represents a fundamental tension between authentic self-expression and social acceptance. \textit{``The fact that it's the same Korean language but a different Korean is what's more confusing''} (P6).

\subsubsection{\textbf{From Fear of Expression to Rights Recognition: Technological Support for Agency Development}}

Participants exhibited a profound fear of asking questions. 
This fear was rooted in past experiences where questioning was perceived as a challenge to authority and could result in punishment. 
Over the course of the test, they demonstrated a gradual progression from fear-based silence to rights-based self-expression. 
This finding suggests that MediBridge's structured guidance approach effectively facilitate this transformation.

\paragraph{\textbf{The Paralyzing Fear of Questioning.}} Participants initially expressed profound anxiety about asking questions, rooted in past experiences where inquiry could lead to punishment. \textit{``In North Korea, asking questions meant you were challenging authority, and that could lead to serious consequences''} (P1). This fear manifested directly in medical settings, where participants would avoid seeking clarification even when confused. \textit{``I would just nod and accept whatever the doctor said, even if I didn't understand. It was safer to stay quiet than risk being seen as difficult''} (P8).

\paragraph{\textbf{Guidelines as ``Permission to Speak.''}} MediBridge's structured guidelines functioned as explicit permission for participants to exercise their rights as patients. \textit{``Seeing 'This question is safe and valid' written there gave me permission I didn't know I needed''} (P11). The systematic nature of the guidance helped participants reframe questioning as collaborative rather than confrontational. \textit{``Having the guidelines written out step by step made me realize that asking about my symptoms isn't challenging the doctor's authority—it's helping them help me''} (P10).

\paragraph{\textbf{Progressive Rights Recognition.}} Through repeated use, participants experienced a fundamental shift in self-perception. \textit{``I started to understand that asking enough questions is actually a right. As a patient, I have the right to understand what's happening to my body''} (P1). This cognitive shift extended beyond medical contexts to broader self-advocacy. \textit{``Using the system changed how I see myself. I realized I'm not just someone who should be grateful for whatever care I get—I'm someone with legitimate needs and rights''} (P8). However, some participants found the structured approach cognitively demanding. \textit{``Sometimes there were so many steps to follow that it felt overwhelming. It reminded me of having to follow complicated rules, which brought back some uncomfortable memories''} (P4).

\subsubsection{\textbf{Structured Approaches as Cognitive Support: Beyond Simple Interface Design}}

Participants' reactions to MediBridge's structured interface challenged conventional assumptions about cognitive load, demonstrating that systematic organization could provide psychological stability for users.

\paragraph{\textbf{The Burden and Benefit of Structured Complexity.}} While participants initially expressed preference for simpler interfaces, their actual usage revealed that MediBridge's structured multi-step process provided unexpected benefits. Although participants noted that the process required more effort, many recognized its value in fostering clarity and confidence. \textit{``Going through each step required more effort, but checking things off one by one helped me feel more organized and confident''} (P7). Rather than feeling overwhelmed by additional elements, participants often found that the systematic organization was more helpful than simplified alternatives. \textit{``Having everything laid out clearly actually made me feel less overwhelmed, even though there were more things to do''} (P15).

\paragraph{\textbf{Choice Fatigue vs. Choice Comfort.}} Participants revealed an unexpected relationship with decision-making that contradicted typical user experience paradigms. Their past experiences in environments with restricted agency affected how they processed options in the interface. \textit{``In North Korea, we didn't have to make many choices about daily things. Here, sometimes having too many options makes me nervous''} (P11). However, this relationship with structured guidance demonstrated inherent tensions between appreciating direction and preserving autonomy. \textit{``I appreciated having the structural guidance, but sometimes it felt like I was being told what to do again, which made me uncomfortable''} (P4).

\paragraph{\textbf{Discovering Hidden Needs Through Systematic Guidance.}} The structured approach proved particularly valuable in helping participants identify concerns they hadn't previously articulated. \textit{``Going through the checklist step by step made me realize things I should be asking about but never thought of''} (P7). This discovery process extended beyond simple information gathering, as participants reported that the systematic prompts revealed gaps in their understanding of patient rights and helped them identify previously unrecognized health communication needs.

\subsubsection{\textbf{The Evaluation Paradox: Trust Without Measurement}}
A notable pattern emerged in participants' relationship with AI assessment system—they embraced the system as a supportive presence while strongly rejecting quantitative evaluations. The findings highlight the importance of considering users' trauma histories in technology design, suggesting that supportive framing and non-evaluative feedback may be essential for vulnerable populations.

\paragraph{\textbf{Trust in the AI `Doctor.'}} Participants readily accepted and trusted MediBridge when it was framed as a supportive medical authority providing a safe practice space. \textit{``When I thought of the AI as a doctor helping me practice, I felt comfortable. I could make mistakes without worrying about real consequences''} (P3). This framing created a judgment-free environment that contrasted sharply with their real-world medical experiences. \textit{``Unlike a real hospital where I worry about being judged or misunderstood, with the AI doctor I could take my time and ask as many questions as I needed''} (P5).

\paragraph{\textbf{Strong Reactions to Scoring.}} In contrast, participants exhibited intense negative responses to quantitative evaluations, often triggering memories of surveillance and assessment in North Korea. \textit{``The moment I saw the score, I felt suffocated. It brought memories of always being watched and evaluated in North Korea''} (P8). These reactions were rooted in past experiences of constant surveillance and assessment. \textit{``Numbers and scores remind me of being measured against impossible standards. In North Korea, you were always receiving grades on your loyalty, your work, everything''} (P2).

\paragraph{\textbf{The Support-Control Distinction.}} Participants distinguished between helpful guidance and controlling evaluation, preferring systems that empowered rather than assessed. \textit{``I want the system to help me, not judge me. When it gives tips or shows what to say, that feels helpful. When it gives a score, that feels like control''} (P10). This distinction was crucial for acceptance and engagement. \textit{``I need to known the system is on my side, not testing me. The difference between 'Here's how you can do better' and 'Here's how you scored' seem small but completely changes how I feel about using it''} (P6).

\subsubsection{\textbf{Addressing Social and Cultural Barriers: The Need for Population-Specific Design}}

Participants' experiences revealed that while generic AI solutions could support basic communication, addressing deeply rooted social and cultural concerns required targeted interventions specifically designed for their unique circumstances.

\paragraph{\textbf{Generic AI's Communication Support but Cultural Gaps.}} Participants appreciated Simple Version's intuitive conversational interface for basic symptom expression. However, they identified critical gaps in addressing their social and cultural concerns. \textit{``The simple version was easy to use but didn't address my main worry—others' perceptions of me''} (P9). Despite improved communication abilities, underlying social anxieties remained unaddressed. \textit{``I could explain symptoms clearly with the simple system, yet still worried whether the doctor would take me seriously''} (P12).

\paragraph{\textbf{Targeted Features for Cultural Concerns.}} MediBridge's culturally informed components specifically addressed participants' social anxieties and identity concerns. The bias-reduction content proved particularly effective in alleviating social fears. \textit{``Reading that 'people focus less on others than you think' significantly reduced my anxiety''} (P11). Participants valued the system's recognition of their specific cultural context. \textit{``The system understood our particular challenges—not just general communication problems—which built my trust''} (P13).

\paragraph{\textbf{The Value of Specificity.}} Participants emphasized the importance of culturally informed design in addressing their particular needs. \textit{``Other apps or simple AI might help with basic communication, but they don't unde}\textit{rstand the constant concern about whether people detect my origins through speech patterns.''} (P14). This suggests that while generic solutions provide foundational support, addressing population-specific challenges requires targeted interventions.

\section{Discussion}

In this section, we discuss five insights from studying NKDs' healthcare experiences and derive design principles for healthcare technologies supporting displaced populations.

\subsection{\textbf{The Linguistic Identity Dilemma: Designing for Oscillation}}

The qualitative interviews identified ``Linguistic Identity Ambivalence'': a simultaneous desire to both preserve and modify one's dialect.
We interpret this contradiction as arising from NKDs' unique positioning as speakers of ``the same yet different'' language, who share grammar but diverging in vocabulary and intonation enough to be identified as North Korean.
Unlike migrants navigating clearly distinct languages, NKDs occupy an ambiguous boundary where their speech is close enough to potentially pass yet distinct enough to risk exposure at any moment.
This creates a persistent tension: complete assimilation feels like losing authentic identity, while maintaining original speech patterns risks being marked as other.
Notably, some participants viewed MediBridge's incomplete dialect recognition as protective rather than problematic, suggesting that technological limitations can paradoxically provide relief by removing the pressure of this constant negotiation.
Similar linguistic ambivalence appears in refugee research~\cite{Reddick2021, McCrocklin2024}, yet prior work typically frames this as a stable preference toward either maintenance or adjustment.
Taken together, these findings suggest a more fluid oscillation driven by context, challenging acculturation theories that assume fixed adaptation strategies~\cite{Berry1997}.

This oscillation challenges existing design approaches.
Current HCI solutions for language barriers typically adopt a single strategy: facilitating user adaptation~\cite{r2_2}, providing translation~\cite{r2_4}, or training providers in cultural competence~\cite{r2_1}.
However, for users oscillating between competing needs, such fixed approaches may prove insufficient.
A system that only facilitates assimilation~\cite{r2_2} risks undermining authenticity, while one that only mediates through translation~\cite{r2_4} risks overlooking the identity dimensions of linguistic choice.
We suggest that effective design should instead prioritize user agency, enabling contextual selection of support type as identity needs shift.
This principle likely extends to other populations navigating contested identities, though for NKDs the stakes of each choice are heightened by expectations of seamless integration with co-ethnic hosts.
Each moment of linguistic disclosure carries the risk of failed belonging, underscoring the need for fine-grained control over when and how adjustments are made.

\subsection{\textbf{Relearning Agency: From Informational Passivity to Active Rights Expression}}

We revealed a transformation in how participants viewed questioning: from dangerous to legitimate.
We interpret this shift as evidence that the primary barrier was not knowledge deficit~\cite{ahn2012} but ``internalized informational passivity,'' where prolonged exposure to authoritarian systems creates lasting barriers to self-advocacy that persist even when information becomes available.
In North Korea, questioning authority figures could lead to serious consequences, teaching citizens that silence is safer than inquiry.
This learned fear does not automatically dissolve upon resettlement; participants described nodding along with doctors even when confused, replicating survival behaviors from their past.
Similar emotional barriers to help-seeking have been identified in other vulnerable populations such as asylum applicants~\cite{r2_5}, suggesting this is not unique to NKDs.
However, the depth of internalization may differ: while other refugees experienced restricted environments for portions of their lives, NKDs were socialized from birth in a system where the very concept of individual rights was absent.
This distinction makes the barrier not just emotional but also conceptual.

MediBridge's effectiveness, then, may stem less from information provision than from functioning as a ``rights rehearsal space.''
The explicit permission statements and structured guidelines did not merely teach participants what to say but validated that speaking was permissible at all.
This distinguishes our approach from general skill-practice tools such as game-based communication trainers~\cite{r2_2}, which help users practice how to speak; our system helped users learn that they have the right to speak.
This principle of providing psychological scaffolding for self-advocacy likely extends to other populations navigating intimidating systems, such as domestic violence survivors seeking legal aid.
For NKDs specifically, designs may need to go further by explicitly introducing patient rights rather than assuming prior familiarity, recognizing that for this population, the foundation for self-advocacy itself may need to be built.

\subsection{\textbf{The Training Wheel Paradox: When Structure Enables Rather Than Constrains}}
The findings revealed an apparent paradox: the simpler Simple Version received higher usability scores (SUS 84.5 vs. 76.5), yet the more structured MediBridge produced significantly higher perceived communication capability scores.
Participants described MediBridge's step-by-step process as more ``work'' but safer and more reassuring.
We interpret this not as a failure of usability but as evidence that cognitive load for this population may stem not from interface complexity but from the psychological burden of autonomous decision-making itself.
For individuals socialized in environments where daily choices were minimized through systems like state job assignments and food rationing, an open-ended interface may feel not liberating but overwhelming.
This aligns with Carroll and Carrithers' Training Wheels principle~\cite{TrainingWheels}, which demonstrated that structured constraints can create safe learning environments by making error states unreachable.
Here, structure functioned not as constraint but as scaffolding that reduced the anxiety of choice.

This finding offers a nuance to conventional design wisdom that universally prioritizes simplification~\cite{cognitiveloadtheory2006, matlin2025digital}.
For users transitioning from institutional or authoritarian contexts, the added structure may paradoxically reduce cognitive load by externalizing decision-making.
The beneficial trade-off we observed (increased interface load exchanged for decreased psychological load) suggests a principle of ``structured simplicity'': using progressive guidance not to constrain but to enable autonomous action.
This likely generalizes to other populations transitioning from high-control environments, such as formerly incarcerated individuals reentering society.
For NKDs specifically, this need may be particularly pronounced given lifelong socialization in systems where individual choice was largely absent, meaning that what appears as unnecessary complexity to other users may function as essential scaffolding for this population.

\subsection{\textbf{The Evaluator Effect: When AI Feedback Triggers Surveillance Memories}}

Participants readily trusted MediBridge when framed as a supportive doctor guiding their practice, yet exhibited strong resistance when the system provided quantitative scores.
The moment numerical evaluation appeared, some participants reported feeling ``suffocated,'' describing memories of constant surveillance in North Korea.
We interpret this pattern as evidence that past experiences of surveillance and judgment shape how users perceive AI's social role.
While AI trust research suggests transparent metrics enhance user confidence~\cite{Zerilli2022, Duarte2023, r2_3}, our findings indicate this approach can backfire when it triggers traumatic associations with evaluation.
This extends the Computers Are Social Actors paradigm~\cite{Nass1994CASA, Gambino2020CASA}: users do not merely perceive AI as social but project specific roles onto it based on experiential background~\cite{r2_7}.
The same feedback mechanism can function as supportive guidance or threatening surveillance depending on how it echoes users' past.

This finding carries particular weight for NKDs, whose experience of evaluation was pervasive and consequential.
Systems like songbun (socio-political classification) and saenghwal chonghwa (weekly self-criticism sessions) meant that being measured against standards was not occasional but constant, not low-stakes but potentially life-altering~\cite{collins2012songbun,collins2018denied}.
This may explain the intensity of participants' reactions to numerical scores, a response that might seem disproportionate without this context.
For trauma-affected populations broadly, this suggests that well-intentioned transparency features can inadvertently exacerbate barriers by triggering the very anxiety they aim to reduce~\cite{Chen2022}.
Design implications point toward prioritizing psychological safety over objective measurement: employing supportive framing rather than numerical scoring, and enabling progress tracking through qualitative guidance rather than quantitative grades.
The goal is empowerment without surveillance, feedback without judgment.

\subsection{\textbf{Designing for Bidirectional Bias: External Stigma and Internalized Doubt}}

The quantitative analysis revealed that the generic Simple Version produced no significant reduction in social and cultural concerns compared to baseline, whereas MediBridge did so significantly.
We interpret this gap through the lens of ``bidirectional bias'': participants simultaneously managed perceived external stigma (fearing judgment from healthcare providers) while also confronting internalized doubt (their own anxieties about whether their concerns were legitimate).
Generic solutions that ignore this dual structure risk addressing only one direction.
Existing approaches such as provider cultural competence training~\cite{r2_1} or translation services~\cite{r2_4} primarily target external barriers, aiming to make the system or communication less hostile.
These are vital interventions, yet our findings suggest such approaches may be insufficient when internal barriers persist.
A participant might encounter a kind doctor and clear communication yet still hesitate to speak, held back by internalized beliefs that their questions are burdensome or their background shameful.

MediBridge's effectiveness in reducing social and cultural concerns appears to stem from its function as a psychological mediator addressing this internal dimension.
The explicit validation that concerns are legitimate and questions are welcome did not change the external environment but changed participants' relationship to it, suggesting that addressing internal psychological barriers represents a design dimension complementary to external systemic ones~\cite{r2_6, r2_7}.
For NKDs specifically, decades of ideological education framing defection as betrayal, combined with societal expectations of seamless integration, may produce a persistent sense of illegitimacy that external kindness alone cannot dissolve.
Designs for this population may therefore need to move beyond cultural adaptation toward explicit psychological validation, affirming not just that users can ask but that they deserve to.

\subsection{\textbf{Ethical Considerations}}

The deployment of AI in healthcare contexts raises ethical questions regarding system limitations and potential risks.
While MediBridge is a communication preparation tool rather than a diagnostic system, AI-generated responses may appear authoritative despite potential inaccuracies~\cite{WHO2025AIGuidance}.
To mitigate this, we explicitly informed users that the AI doctor is for practice purposes only and does not provide medical advice, diagnosis, or treatment.
The prototype was designed to complement actual clinical encounters, with the ``Helper Note'' encouraging users to seek real medical consultation.
Future deployment should incorporate mechanisms to detect and intervene if users exhibit over-reliance on AI practice or delay seeking actual healthcare.

Research with displaced populations demands heightened attention to participant protection and data privacy~\cite{Seagle2020, birman2006ethical}.
North Korean defectors face unique vulnerabilities: identity disclosure risks harm to family members remaining in North Korea, social stigmatization in South Korea, and psychological re-traumatization from surveillance-related experiences~\cite{Noh2024}.
Following trauma-informed computing principles~\cite{Chen2022}, our research incorporated anonymous participant identifiers, researcher team composition with community ties to establish psychological safety, and flexible consent protocols allowing withdrawal at any point.
For future deployment, voice recordings, conversation logs, and dialect analysis data constitute highly sensitive information for this population.
Such systems require robust encryption, explicit user control over data retention, and strict policies preventing identity-compromising data sharing.

\section{Limitations and Future Work}
This study has several limitations that reflect our methodological choices and research focus.
First, we deliberately focused on North Korean defectors from the Seoul metropolitan area to ensure depth of analysis and cultural homogeneity, though this limits generalizability to other regions or populations from different authoritarian backgrounds.
Second, we conducted evaluations in controlled settings to isolate system's core effects from confounding variables, leaving real-world effectiveness unexamined.
Third, we examined short-term usage effects to understand initial user responses and system acceptance, as prolonged exposure studies would require different ethical considerations and participant commitment that were beyond the scope of this initial investigation.
Finally, while we introduce theoretical concepts such as ``Linguistic Identity Ambivalence'' and ``Internalized Informational Passivity'' based on our empirical findings, these require validation across different populations to establish their broader applicability in HCI research.

Future research remains to build upon these foundational findings through several key directions. Conducting in-the-wild deployments in actual clinical settings would assess the practical effectiveness of MediBridge's features and its impact on real patient-provider interactions and health outcomes. Longitudinal studies tracking users over months or years could reveal how trauma recovery, linguistic adaptation, and technology acceptance evolve over time, providing insights into optimal intervention timing and duration. Expanding research to include NKDs at various adaptation stages and from different regions, as well as refugees from other authoritarian backgrounds (e.g., Myanmar or Cuba), would strengthen the generalizability of findings and validate the theoretical frameworks we propose. Additionally, developing and testing trauma-informed AI design principles more systematically could establish evidence-based guidelines for creating supportive technologies for vulnerable populations. Finally, investigating alternative approaches to progress tracking and motivation that avoid quantitative evaluation could address the ``evaluation allergy'' phenomenon we identified, potentially leading to more effective and psychologically safe systems for trauma-affected users.

\section{Conclusion}

We investigated North Korean defectors' healthcare challenges, identifying clinical consultation as a critical barrier.
We designed and evaluated MediBridge, an AI-powered system that enables rehearsal with a simulated doctor and generates personalized ``Helper Notes'' for actual visits, providing evidence of improved perceived communication capability across multiple dimensions.
Beyond this artifact, the primary contribution of this study lies in studying this unique population to surface design considerations often invisible in general refugee research, including navigating linguistic identity oscillation in ``same yet different'' language contexts and providing rights rehearsal for users from authority-sensitive backgrounds.
We offer these design principles for designing healthcare technologies that serve displaced populations at the intersection of systemic unfamiliarity and psychological recovery.

\begin{acks}
This work was supported by the New Faculty Startup Fund from Seoul National University. 
This work was also supported by the National Research Foundation of Korea(NRF) grant funded by the Korea government(MSIT) (No. RS-2024-00407105).
This study was also supported by the Unification and Peace Research Support Program of the 
Institute for Peace and Unification Studies (IPUS) at Seoul National University in 2025.
This work was partly supported by Institute of Information \& Communications Technology Planning \& Evaluation (IITP) grant fundedby the Korea government(MSIT) [NO.RS-2021-II211343, Artificial Intelligence Graduate School Program (Seoul National University)] and Basic Science Research Program through the National Research Foundation of Korea (NRF) funded by the Ministry of Education(No. RS-2025-25421701)
\end{acks}

\bibliographystyle{ACM-Reference-Format}
\bibliography{ref}

\end{document}